\documentclass[twocolumn]{aastex62}

\usepackage{soul}
\received{January 1, 2018}
\revised{January 7, 2018}
\accepted{\today}
\submitjournal{ApJ}

\shorttitle{UV radiation and/or X-ray effects on the Sgr B2 cloud}
\shortauthors{Armijos-Abenda\~no et al.}

\begin{document}

\title{On the effects of UV photons/X-rays on the chemistry of the Sgr B2 cloud} 

\correspondingauthor{J. Armijos-Abenda\~no}
\email{jairo.armijos@epn.edu.ec}

\author{J. Armijos-Abenda\~no}
\affil{Observatorio Astron\'omico de Quito, Escuela Polit\'ecnica Nacional, Interior del Parque La Alameda, 170136, Quito, Ecuador}

\author{J. Mart\'in-Pintado}
\affiliation{Centro de Astrobiolog\'ia (CSIC, INTA), Ctra a Ajalvir, km 4, 28850, Torrej\'on de Ardoz, Madrid, Spain}

\author{E. L\'opez}
\affiliation{Observatorio Astron\'omico de Quito, Escuela Polit\'ecnica Nacional, Interior del Parque La Alameda, 170136, Quito, Ecuador}

\author{M. Llerena}
\affiliation{Observatorio Astron\'omico de Quito, Escuela Polit\'ecnica Nacional, Interior del Parque La Alameda, 170136, Quito, Ecuador}
\affiliation{Departamento de F\'isica y Astronom\'ia, Universidad de La Serena, Av. Juan Cisternas 1200 Norte, La Serena, Chile}

\author{N. Harada}
\affiliation{Academia Sinica Institute of Astronomy and Astrophysics, P.O. Box 23-141, Taipei 10617, Taiwan}

\author{M. A. Requena-Torres}
\affiliation{Department of Astronomy, University of Maryland, College Park, MD 20742, USA}

\author{S. Mart\'in}
\affiliation{European Southern Observatory, Alonso de C\'ordova 3107, Vitacura, Santiago, Chile}
\affiliation{Joint ALMA Observatory, Alonso de C\'ordova 3107, Vitacura, Santiago, Chile}

\author{V. M. Rivilla}
\affiliation{INAF/Osservatorio Astrofisico di Arcetri, Largo Enrico Fermi 5, I-50125, Florence, Italy}

\author{D. Riquelme}
\affiliation{Max-Planck-Institut f\"ur Radioastronomie, Auf dem H\"ugel 69, 53121 Bonn, Germany}

\author{F. Aldas}
\affiliation{Observatorio Astron\'omico de Quito, Escuela Polit\'ecnica Nacional, Interior del Parque La Alameda, 170136, Quito, Ecuador}







\begin{abstract}
The lines of HOC$^+$, HCO and CO$^+$ are considered good tracers of photon-dominated regions (PDRs) and X-ray dominated regions. We study these tracers towards regions of the Sgr B2 cloud selected to be affected by different heating mechanisms. We find the lowest values of the column density ratios of HCO$^+$ versus HOC$^+$, HCO and CO$^+$ in dense HII gas, where UV photons dominate the heating and chemistry of gas. HOC$^+$, HCO and CO$^+$ abundances and the above ratios are compared with those of chemical modeling, finding that high temperature chemistry, a cosmic-ray ionization rate of 10$^{-16}$ s$^{-1}$ and timescales $>$10$^{5.0}$ years explain well the HOC$^+$ abundances in quiescent Sgr B2 regions, while shocks are also needed to explain the highest HCO abundances derived for these regions. CO$^+$ is mainly formed in PDRs since the highest CO$^+$ abundances of $\sim$(6-10)$\times$10$^{-10}$ are found in HII regions with electron densities $>$540 cm$^{-3}$ and that CO$^+$ emission is undetected in quiescent gas. Between the ratios, the HCO$^+$/HCO ratio is sensitive to the electron density as it shows different values in dense and diffuse HII regions. We compare SiO J=2-1 emission maps of Sgr B2 with X-ray maps from 2004 and 2012. One known spot shown on the 2012 X-ray map is likely associated with molecular gas at velocities of 15-25 km s$^{-1}$. We also derive the X-ray ionization rate of $\sim$10$^{-19}$ s$^{-1}$ for Sgr B2 regions pervaded by X-rays in 2004, which is quite low to affect the chemistry of the molecular gas.

\end{abstract}



\section{Introduction}\label{sec:intro}
One of the key issues in extragalactic astrophysics is the search for molecular
tracers of the heating mechanisms in nearby galactic nuclei \citep{Usero04,Fuente06,Martin09a}. These tracers could be used
as probes of the heating sources to disentangle the type of activity in more distant galaxies. As a result of years of investigation,
several molecular tracers have been proposed to discriminate between
the main heating sources in active galaxies \citep{Martin09b,Martin15}.
Whereas galaxies hosting active galactic nuclei (AGNs) are affected by hard X-rays arising from the accretion disks of supermassive black holes, the chemistry and heating of starburst galaxies are dominated by shocks, UV radiation, and/or cosmic rays
depending on the evolutionary stage of their nuclei \citep{Aladro11}. However, this picture is not simple since extragalactic observations, even in the case of the closest galaxies, do not have the necessary spatial resolution to isolate the effects of the different heating sources. Located
at only 7.9 kpc \citep{Boehle16}, the Galactic Center (GC) offers an unique opportunity to study in detail a galactic nucleus. This proximity allows to discriminate regions according to the main heating source affecting the molecular gas.

The emission lines of HCO and HOC$^+$ are considered to be good tracers of both PDR (Photon-dominated region) and XDR
(X-ray Dominated Region) chemistries \citep{Ziurys95,Apponi99,Usero04, Martin09a}. The abundance of CO$^+$ is thought to be enhanced in regions affected by strong UV and X-ray radiation fields \citep{Fuente06,Spaans07}. High abundances of HCO and HOC$^+$ have been found in NGC 1068, which is considered a prototypical extragalactic XDR \citep{Usero04}. HCO, HOC$^+$ and CO$^+$ are abundant in the starburst galaxies M 82 and NGC 253 \citep{Fuente06,Martin09a,Aladro15}. 

The HOC$^+$(1-0) line was detected towards the circumnuclear disk of the NGC 1808 galaxy, where the chemistry of HOC$^+$ may be affected by PDRs or XDRs \citep{Salak18}.
HOC$^+$(3-2) shows emission from a very compact region close to the nucleus of the ULIRG (ultra-luminous infrared galaxy) Mrk 273 unlike the 1-0 and 3-2 transitions of HCN, HNC, and HCO$^+$ that show a more extended emission, suggesting that the chemistry of HOC$^+$ is different from that of the other molecules \citep{Aladro18}.
HOC$^+$ emission traces the molecular cloud layers most exposed to the UV radiation in the Orion Bar \citep{Goicoechea17}.

CO$^+$ emission is tracing the layers between the HII region and the molecular gas in the Mon R2 star forming region \citep{Trevino16}. CO$^+$ formation is expected in dense regions affected by high UV fields \citep{Trevino16}.

The 6.4 keV Fe K$\alpha$ line is considered to be an excellent tracer of XDRs as this line arises from fluorescence when high energy X-rays and/or particles ($>$7.1 keV) interact with neutral or partially ionized iron atoms \citep{Amo09}. 
The X-ray emission towards the GC has been studied by \cite{Terrier18} using the 6.4 keV Fe K$\alpha$ line, finding that most regions they studied in the GC show bright X-ray emission in 2000-2001, which significantly decrease by 2012.
\cite{Kawamuro19} used the 6.4 keV Fe K$\alpha$ line to map the X-ray irradiated gas towards the $\sim$100 pc central region of the Circinus galaxy.

\begin{figure}
\caption{\textbf{Top panel:} the thirteen positions studied in this paper are indicated as circles with the size of the 30 meter IRAM telescope beam of 28\arcsec at 86 GHz on the Sgr B2 radio continuum map at 20 cm \citep{Yusef04}. Positions 1 and 2 trace dense HII regions towards Sgr B2M (it is the origin of the offsets in arcsec) and Sgr B2N, respectively, positions 3, 5, 7-9 trace diffuse HII regions, while positions 4, 6, 10-13 trace regions of quiescent gas (see Section \ref{ionized_hydrogen}). The gas in positions 4, 6, 10-12 are pervaded by X-rays (see middle panel), however the effects of X-rays on the gas in these five positions are negligible (see Section \ref{X_ray_effects}), thus these regions are considered quiescent.
\textbf{Middle and bottom panels:} same as in the top panel, but the positions are overlapped on the 6.3-6.5 keV color maps observed in 2004 (middle panel) and 2012 (bottom panel), where the G0.66-0.03, G0.74-0.10 and G0.66-0.13 sources studied in \cite{Terrier18} are highlighted with big white ellipses. We called the X-ray source observed in 2004 as G0.69-0.11. Contours show the SiO J=2-1 emission integrated over the velocity ranges of [65-75] and [15-25] km s$^{-1}$ in the middle and bottom panels, respectively (see Section \ref{Xrays}). Contour levels start at 1 K km s$^{-1}$ (3$\sigma$) and increase in 2.2 K km s$^{-1}$ steps in both velocity ranges.}
\includegraphics[width=0.41\textwidth]{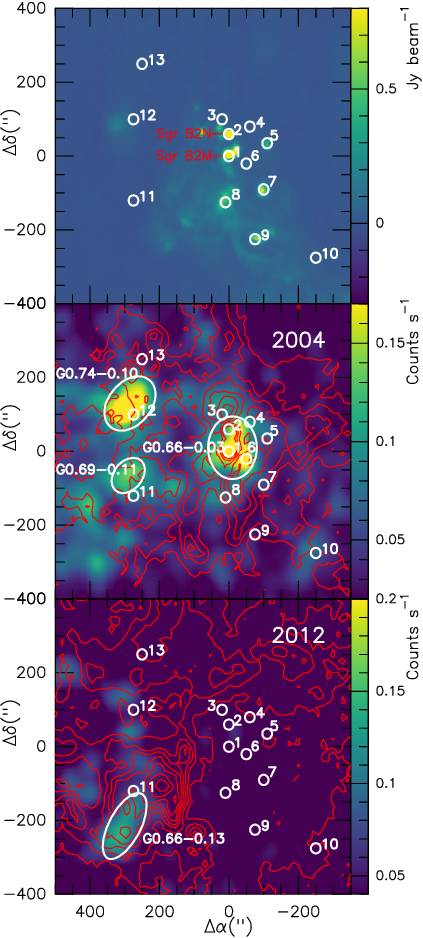}
\label{Pos_figure}
\end{figure}

So far, it is unclear to what extent the proposed PDR and/or XDR tracers, HCO, HOC$^+$ and CO$^+$, may be used to disentangle the PDR and XDR environments in galactic nuclei.
In this paper, we will try to address the above ambiguity by studying these molecular tracers in spatially resolved regions affected by X-rays and/or UV radiation in the Sgr B2 cloud. We have chosen thirteen positions for our study. 
We give in Table \ref{Pos_SgrB2} the heating mechanisms that affect the sources included in this paper. Positions 1, 2, 5, 7-8 were selected towards 20 cm emission peaks shown on the top panel of Figure~\ref{Pos_figure}, therefore the chemistry and heating of the gas in these regions would be dominated by the UV radiation. In Section \ref{ionized_hydrogen}, we will see that these regions can be classified into HII regions of diffuse or dense gas, including position 3 to the regions of diffuse HII gas (see Section \ref{ionized_hydrogen}). As seen on middle panel of Figure~\ref{Pos_figure}, the gas in positions 1-8 and 10-12 are pervaded by X-ray emission above 6$\sigma$, while 9, and 13 are not. The possible effects of X-rays on the chemistry of several of these Sgr B2 positions will be studied and discussed in Sections \ref{Xrays} and \ref{X_ray_effects}, respectively.

\begin{deluxetable}{ccc}
\tablecaption{Positions observed in Sgr B2\label{Pos_SgrB2}}
\tablehead{
Pos. & Heating & $\Delta\alpha$\tablenotemark{a},$\Delta\delta$\tablenotemark{a} \\
     & mechanism &  (\arcsec,\arcsec)}
\startdata
1 & UV photons and/or X-rays & (0, 0) \\
2 & UV photons and/or X-rays & (0,60)  \\
3 & X-rays & (20,100) \\
4 & X-rays & (-60,80) \\
5 & UV photons and/or X-rays & (-110,35) \\
6 & X-rays & (-50,-20) \\
7 & UV photons and/or X-rays & (-100,-90) \\
8 & UV photons and/or X-rays & (10,-125) \\
9 & UV photons & (-75,-225) \\
10 & X-rays & (-250,-275) \\
11 & X-rays & (275,-120) \\
12 & X-rays & (275,100) \\
13 & Quiescent & (250,250) \\
\enddata
\tablenotetext{a}{Offsets are relative to the position of the massive hot core Sgr B2M ($\alpha_{J2000}$=17:47:20.40 and $\delta_{J2000}$=-28:23:07.25).}
\end{deluxetable}

\begin{figure}
\caption{{\bf Top panel}: Sgr B2 emission distribution of the HCO 1$_{0,1}$-0$_{0,0}$ (J=3/2-1/2, F=2-1) line integrated over the velocity range of 40-80 km s$^{-1}$. We show positions 1-10 studied in this paper as circles with the size of the 30 meter IRAM telescope beam of 28\arcsec at 86 GHz. Positions 11-13 are outside the region mapped in the HCO(1$_{0,1}$-0$_{0,0}$) line. Position 1 corresponds to Sgr B2M and it is the origin of the offsets in arcsec.
Contours in red start at 1.6 K km s$^{-1}$ (2$\sigma$) and increase in 0.6 K km s$^{-1}$ steps. The wedge at the right side of the panel indicates the intensity scale in K km s$^{-1}$.
\mbox{{\bf Bottom panel}:} same as in the top panel, but the positions are drawn on the emission of the H$^{13}$CO$^+$ J=1-0 line integrated over the same velocity range as above. Contours start at 2.7 K km s$^{-1}$ (3$\sigma$), increasing in 2.3 K km s$^{-1}$ steps. In this panel, positions 11-13 are drawn as the mapped region is larger than that shown in the top panel.}
\centering
\includegraphics[width=0.4\textwidth]{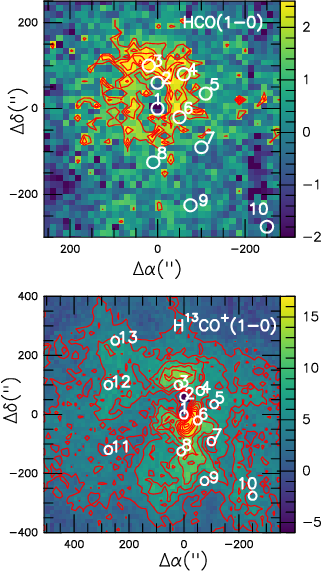}
\label{hcoemission}
\end{figure}

\begin{figure}
\caption{Fe K$\alpha$ line fluxes measured towards the thirteen positions of Sgr B2. Red and black symbols show the 2004 and 2012 values, respectively. Triangles in red and black show upper limits on the Fe K$\alpha$ line flux.}
\centering
\includegraphics[width=0.5\textwidth]{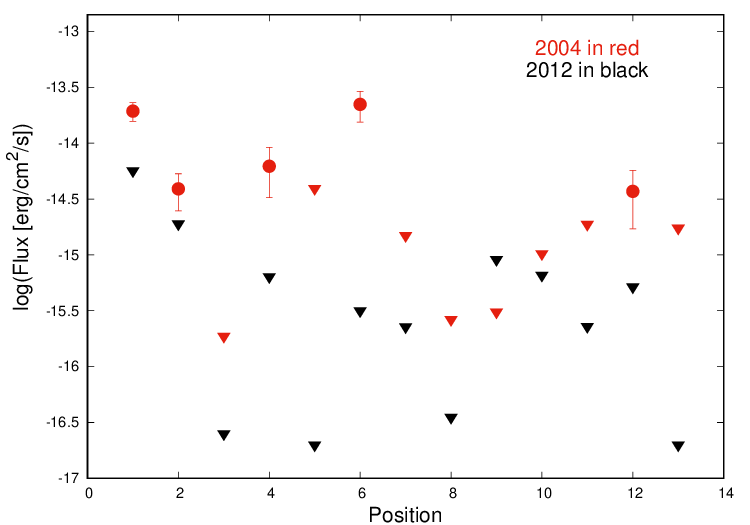}
\label{Fe6.4_flux}
\end{figure}

\begin{figure*}
\caption{X-ray spectra observed towards four positions (1, 3, 9 and 12) of Sgr B2. The red and black crosses show the XMM-Newton observations taken in 2004 with each EPIC-MOS camera, while the histograms represent the best-fit models. The lower panels in each box show the residuals normalized to the errors.\label{Xray_spectrum}}
\gridline{\rotatefig{-90}{position1q.eps}{0.3\textwidth}{}
          \rotatefig{-90}{position3q.eps}{0.3\textwidth}{}
          }
\gridline{\rotatefig{-90}{position9q.eps}{0.3\textwidth}{}
          \rotatefig{-90}{position12q.eps}{0.3\textwidth}{}
         }
\end{figure*}

\begin{deluxetable}{crcc}
\tablecaption{Observed molecular and radio recombination lines\label{Freq_lines}}
\tablehead{
Lines & $\nu$ & E$_{\rm u}$ & HPBW\\
     &  (GHz) & (K) & (\arcsec)}
\startdata
HCO(1$_{0,1}$-0$_{0,0}$)\tablenotemark{a} & 86.671 & 4.18 & 28.3\\
HCO(3$_{0,3}$-2$_{0,2}$)\tablenotemark{b} & 260.060 & 24.98 & 9.5\\
HOC$^+$(1-0) &  89.487 &  4.29  & 27.6\\
HOC$^+$(3-2) & 268.451 & 25.77 & 9.2\\
CO$^+$(J=2-1, 3/2-3/2) &  235.380 & 19.96 & 10.5\\
CO$^+$(J=2-1, 5/2-3/2) &  236.062 & 17.00 & 10.4\\
HC$^{18}$O$^+$(1-0) & 85.162 & 4.09 & 28.9\\
HC$^{18}$O$^+$(3-2) & 255.480 & 24.52 & 9.6\\
C$^{18}$O(1-0) & 109.782 & 5.27 & 22.4\\
SiO(2-1) & 86.847 & 6.25 & 28.3\\
H$^{13}$CO$^+$(1-0) & 86.754 & 4.16 & 28.4\\
H42$\alpha$    &  85.688 & \nodata & 28.7\\
\enddata
\tablenotetext{a}{Formed by four hyperfine lines but only the frequency of the most intense of the four lines is given.}
\tablenotetext{b}{Formed by six hyperfine lines but only five of the six lines were observed and the frequency of the most intense of the five lines is given.}
\end{deluxetable}

\section{Observations and data reduction}
Our study was performed by using data obtained with the IRAM 30 meter telescope\footnote{IRAM is supported by INSU/CNRS (France), MPG (Germany) and IGN (Spain).} at Pico Veleta (Spain) and archive data taken with the XMM-Newton space telescope.

\subsection{IRAM 30 meter observations}\label{observa}
With the IRAM 30 meter telescope we observed the HCO, HOC$^+$ and CO$^+$ lines (listed in Table \ref{Freq_lines}) on December 2014 towards thirteen selected positions (Program 137-14; PI: Armijos-Abenda\~no), whose offsets relative to the massive hot core Sgr B2M position (with $\alpha_{J2000}$=17:47:20.40 and $\delta_{J2000}$=-28:23:07.25) are indicated in Table \ref{Pos_SgrB2} and Figure~\ref{Pos_figure}. 
The EMIR receiver (Eight MIxer Receiver) was used during our observations, tuning horizontal and vertical polarizations in the E090 and E230 bands, while the FTS (Fast Fourier Transform Spectrometer) and WILMA (Wideband Line Multiple Autocorrelator) were used as backends.
The FTS backend worked at a resolution of 200 kHz.
We used position switching as observing mode with the emission-free reference position $\alpha_{J2000}$=17:46:23.0 and $\delta_{J2000}$=-28:16:37.3, which was selected from the CS maps by \cite{Bally87}.
An on-source integration time of 30 min was used for each position.
Spectra were calibrated by using ambient and cold temperature loads, which provided a calibration accuracy around 10\%.
The spatial resolution provided by the IRAM telescope is listed in Table \ref{Freq_lines}, which allowed us to separate spatially PDRs, XDRs and quiescent regions (see Figure~\ref{Pos_figure}).
The output spectra obtained from the IRAM 30 meter telescope are calibrated in the antenna temperature scale (T$_{\rm a}^*$), which we converted to main beam temperature (T$_{\rm mb}$) using the relation T$_{\rm mb}$=(F$_{\rm eff}$/B$_{\rm eff}$)T$_{\rm a}^*$, where F$_{\rm eff}$ is the forward efficiency and B$_{\rm eff}$ is the main-beam efficiency.
The B$_{\rm eff}$ values were calculated using the Ruze formula B$_{\rm eff}$=1.2$\epsilon\,e^{-(4\pi\,R\sigma/\lambda)^2}$, where $\epsilon$=0.69 and $R\sigma$=0.07. We calculated B$_{\rm eff}$ values of 0.78-0.45 for the observed frequencies of 85-268 GHz. The F$_{\rm eff}$ values were taken from the online documentation of the IRAM 30 meter telescope\footnote{http://www.iram.es/IRAMES/mainWiki/Iram30mEfficiencies}.

The calibrated data observed in 2014 were imported in the MADCUBA software \citep{Martin19} for further processing. With this software the baseline (of order 0-2) subtraction and spectrum averaging were applied. Then the spectra were smoothed to a velocity resolution of $\sim$5 km s$^{-1}$ appropriate to resolve the typical linewidths observed in the GC. The rms noise level of the spectra is $\sim$10-15 mK in the T$_{\rm mb}$ scale.

\begin{figure}
\caption{Spectra of the H42$\alpha$ RRL observed towards thirteen positions of Sgr B2 (shown in Fig.~\ref{Pos_figure}). The positions are labeled with numbers on the left top corner of each panel. Velocity components observed in the C$^{18}$O J=1-0 lines (see Section \ref{C18O_section}) are indicated with dashed lines for comparison between the ionized gas velocities and those of molecular gas. Positions 1 and 2 trace regions of dense HII gas with electron densities of 1200-1700 cm$^{-3}$, while positions 3, 5, 7-9 trace regions of diffuse HII gas with electron densities of 280-550 cm$^{-3}$ (see Section \ref{ionized_hydrogen}). Positions 1-2, 5, 7-9 were selected from 20 cm continuum emission peaks shown in Figure~\ref{Pos_figure}. The H42$\alpha$ line intensity in positions 1 and 2 are divided by 9 for better visualization.}
\includegraphics[width=0.49\textwidth]{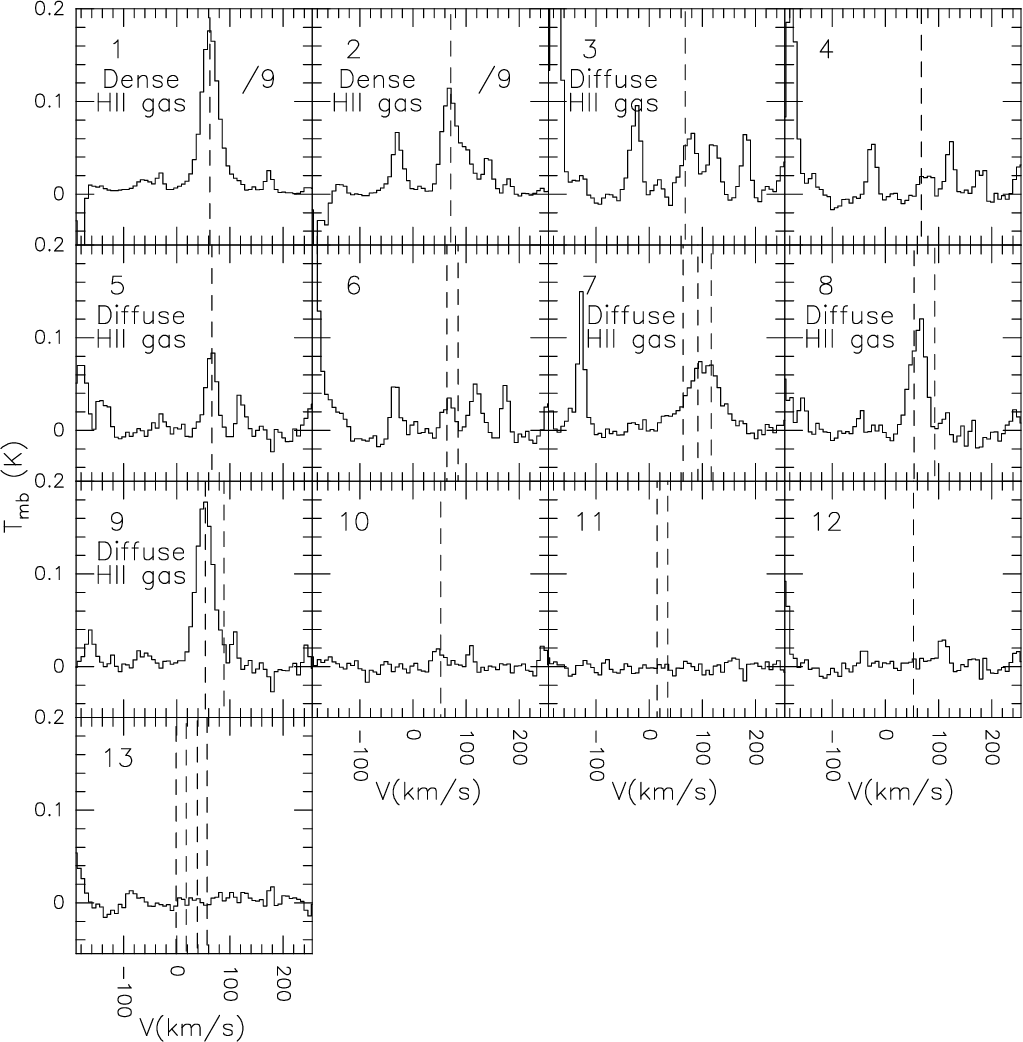}
\label{h42a_figure}
\end{figure}

\begin{figure*}
\caption{HOC$^+$ J=1-0 (top) and 3-2 (bottom) transitions observed towards thirteen positions of Sgr B2 (shown in Fig.~\ref{Pos_figure}). The positions are labeled with numbers on the left top corner of each panel. The LTE best fits to the lines obtained with MADCUBA are shown with red lines. Multiple velocity components shown with dashed lines are considered in the line fitting of positions 6--9, 11 and 13 (see Section \ref{hocp_section}). In positions 1 and 2, the spectra of the HOC$^+$ transitions 1-0 and 3-2 are divided by 2 and 8, respectively, for better visualization. The HOC$^+$ 3-2 line is blended with hyperfine lines of $^{33}$SO$_2$(7$_{2,6}$-6$_{1,5}$) in \mbox{position 1} (see Section \ref{hocp_section}).}
\centering
\includegraphics[width=1\textwidth]{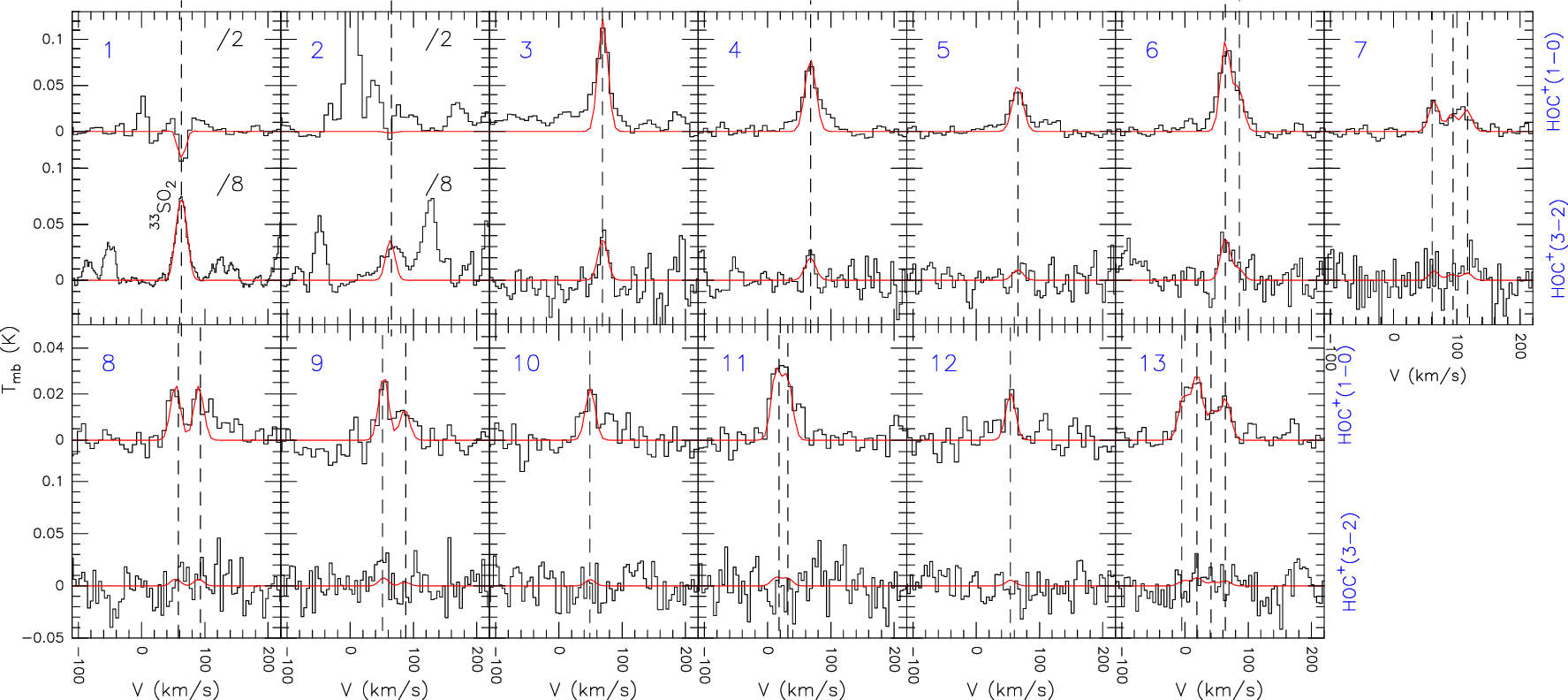}
\label{hocp_figure}
\end{figure*}

\begin{deluxetable}{cccc}
\tablecaption{Best-fit results of the 2004 X-ray spectrum modeling\label{Xray_table}}
\tablehead{
Pos. & EW\tablenotemark{a} & log(L$_{\rm X}$\tablenotemark{b}) & $\chi^2$/d.o.f\tablenotemark{c}\\
     & (keV) & (erg s$^{-1}$) & }
\startdata
1 & 0.63 ($<$0.84) & 37.1$^{+0.1}_{-0.1}$ & 55.38/49 \\
3 & $<$0.49        & 35.9$^{+0.3}_{-1.6}$ & 10.97/10 \\
9 & $<$0.71        & 36.0$^{+0.2}_{-0.5}$ & 17.38/10 \\
12& 0.74 ($<$1.42)&  36.6$^{+0.2}_{-0.3}$ & 12.87/14 \\
\enddata
\tablenotetext{a}{The equivalent width (EW) of the 6.4 keV Fe line.}
\tablenotetext{b}{Absorption-corrected luminosity in the 2-10 keV range.}
\tablenotetext{c}{Degrees of freedom (d.o.f).}
\end{deluxetable}

\begin{longrotatetable}
\begin{deluxetable*}{ccccc|cccc|cccc}
\tablecaption{Derived parameters of the studied molecules\label{Tab_column}}
\tablehead{
Pos. & \multicolumn{4}{c}{HOC$^+$} & \multicolumn{4}{c}{HCO} & \multicolumn{4}{c}{CO$^+$}\\
 & T$_{\rm ex}$ & V$_{\rm LSR}$ & FWHM & N$_{\rm tot}$ & T$_{\rm ex}$ & V$_{\rm LSR}$ & FWHM & N$_{\rm tot}$ & T$_{\rm ex}$ & V$_{\rm LSR}$ & FWHM & N$_{\rm tot}$\\
 & (K) & (km s$^{-1}$) & (km s$^{-1}$) & ($\times$10$^{12}$ cm$^{-2}$) & (K) & (km s$^{-1}$) & (km s$^{-1}$) & ($\times$10$^{13}$ cm$^{-2}$) & (K) & (km s$^{-1}$) & (km s$^{-1}$) & ($\times$10$^{13}$ cm$^{-2}$)
}
\startdata
1 & 4\tablenotemark{a} & 63\tablenotemark{a} & 15\tablenotemark{a} & 10.0$\pm$0.5 &5\tablenotemark{a} & 63\tablenotemark{a} & 13.3$\pm$2.2 & 30.0$\pm$4.5 & 5\tablenotemark{a} & 62\tablenotemark{a} & 14\tablenotemark{a} & 19.0$\pm$8.1\\
\hline
2 & 5\tablenotemark{a} & 65\tablenotemark{a} & 15\tablenotemark{a} & 15.0$\pm$1.0 &5\tablenotemark{a} & 65\tablenotemark{a} & 15\tablenotemark{a} & 20.0$\pm$3.5  & 5\tablenotemark{a} & 65\tablenotemark{a} & 15\tablenotemark{a} & 13.0$\pm$4.1\\
\hline
3 & 5.8$\pm$0.4 & 69.5$\pm$0.9 & 20\tablenotemark{a} & 2.6$\pm$0.2 & 4.4$\pm$0.7 & 70.1$\pm$1.8 & 23.1$\pm$4.3 & 10.0$\pm$2.1 & 5\tablenotemark{a} & 71\tablenotemark{a} & 18\tablenotemark{a} & 3.7$\pm$1.8\\
\hline
4 & 5.6$\pm$0.5 & 67.6$\pm$1.0 & 20\tablenotemark{a} & 1.7$\pm$0.2 & 4.4$\pm$0.5 & 70.1$\pm$0.7 & 18.7$\pm$1.6 & 5.9$\pm$0.6 & 5\tablenotemark{b} & \nodata & 18\tablenotemark{b} & $<$4.2\tablenotemark{c}\\
\hline
5 & 5.2$\pm$0.7 & 66.1$\pm$1.3 & 20\tablenotemark{a} & 1.2$\pm$0.2 & 4.5$\pm$0.8 & 67.0$\pm$1.0 & 16.6$\pm$2.2 & 2.3$\pm$0.3 & 5\tablenotemark{b} &\nodata & 18\tablenotemark{b} & $<$3.4\tablenotemark{c}\\
\hline
6 & 6.3$\pm$0.4 & 63.7$\pm$1.0 & 20\tablenotemark{a} & 2.0$\pm$0.2 & 5\tablenotemark{a} & 64$\pm$1.0 & 18\tablenotemark{a} & 3.7$\pm$0.5 & 5\tablenotemark{b} & \nodata & 18\tablenotemark{b} &  $<$3.4\tablenotemark{c}\\
 & 5.3$\pm$1.1 & 86\tablenotemark{a} & 20\tablenotemark{a} & 1.0$\pm$0.2 & 5\tablenotemark{a} & 85\tablenotemark{a} & 18\tablenotemark{a} & 1.4$\pm$0.6 & 5\tablenotemark{b} & \nodata & 18\tablenotemark{b} & $<$3.4\tablenotemark{c}\\
\hline
7 & 5.8$\pm$1.2 & 61.3$\pm$2.4 & 18\tablenotemark{a} & 0.7$\pm$0.2 & 6.1$\pm$1.4 & 61.2$\pm$2.9 & 18\tablenotemark{a} & 1.2$\pm$0.3 & 5\tablenotemark{b} & \nodata & 18\tablenotemark{b} & $<$3.0\tablenotemark{c}\\
 & 4.6$\pm$2.7 & 94\tablenotemark{a} & 18\tablenotemark{a} & 0.4$\pm$0.1 & 5\tablenotemark{b} & \nodata & 18\tablenotemark{b} & $<$1.1\tablenotemark{c}  & 5\tablenotemark{b} & \nodata & 18\tablenotemark{b} & $<$3.0\tablenotemark{c}\\
 & 6\tablenotemark{a} & 117\tablenotemark{a} & 18\tablenotemark{a} & 0.5$\pm$0.1 & 5\tablenotemark{b} & \nodata & 18\tablenotemark{b} & $<$1.1\tablenotemark{c} & 5\tablenotemark{b} & \nodata &18\tablenotemark{b} & $<$3.0\tablenotemark{c}\\
\hline
8 & 6\tablenotemark{a} & 57.8$\pm$3.4 & 18\tablenotemark{a} & 0.5$\pm$0.1 & 5\tablenotemark{a} &  54.0$\pm$1.8 & 18.5$\pm$4.0 & 3.6$\pm$0.7 & 5\tablenotemark{b} & \nodata & 18\tablenotemark{b} & $<$3.2\tablenotemark{c}\\
  & 7.3$\pm$2.2 & 93.3$\pm$4.4 & 18\tablenotemark{a} & 0.4$\pm$0.2 &5\tablenotemark{b} & \nodata & 18\tablenotemark{b} & $<$0.8\tablenotemark{c} & 5\tablenotemark{b} & \nodata & 18\tablenotemark{b} & $<$3.2\tablenotemark{c}\\
\hline
9  &  9.9$\pm$2.0 & 50.7$\pm$1.9 & 19.4$\pm$4.6 & 0.5$\pm$0.1 & 5.3$\pm$1.9 & 54.6$\pm$3.0 & 15.9$\pm$7.1 & 1.8$\pm$0.7 & 5\tablenotemark{a} & 49.0$\pm$1.4 & 18\tablenotemark{a} &  4.3$\pm$0.6\\
  & 6\tablenotemark{a} & 88\tablenotemark{a} & 18\tablenotemark{a} & 0.2$\pm$0.1 & 5\tablenotemark{b} & \nodata & 18\tablenotemark{b} & $<$0.5\tablenotemark{c} & 5\tablenotemark{b} & \nodata & 18\tablenotemark{b} & $<$3.0\tablenotemark{c}\\
\hline
10 & 6\tablenotemark{a} & 48.9$\pm$4.4 & 18\tablenotemark{a} & 0.4$\pm$0.2 & 5\tablenotemark{a} & 55.2$\pm$1.8 & 12.4$\pm$4.7 & 0.8$\pm$0.3 & 5\tablenotemark{b} & \nodata & 18\tablenotemark{b} & $<$3.8\tablenotemark{c}\\
\hline
11 & 6\tablenotemark{a} & 17.8$\pm$4.1 & 18\tablenotemark{a} & 0.6$\pm$0.2 & 7.9$\pm$1.7 & 4.3$\pm$3.1 & 18\tablenotemark{a} & 0.8$\pm$0.2 & 5\tablenotemark{b} & \nodata & 18\tablenotemark{b} & $<$3.0\tablenotemark{c}\\
  & 6\tablenotemark{a} & 32\tablenotemark{a} & 18\tablenotemark{a} & 0.3$\pm$0.2 & 5\tablenotemark{a} & 28\tablenotemark{a}& 18\tablenotemark{a} & 0.5$\pm$0.3 & 5\tablenotemark{b} & \nodata & 18\tablenotemark{b} & $<$3.0\tablenotemark{c}\\
  \hline
12 & 6\tablenotemark{a} & 54.4$\pm$2.7 & 18\tablenotemark{a} & 0.5$\pm$0.1 & 5.9$\pm$0.7 & 58.4$\pm$1.0 & 12.9$\pm$2.2 & 1.8$\pm$0.3 & 5\tablenotemark{b} & \nodata & 18\tablenotemark{b} & $<$2.3\tablenotemark{c}\\
\hline
13 & 6\tablenotemark{a} & -4.9$\pm$5.2 & 18\tablenotemark{a} & 0.3$\pm$0.1 & 5\tablenotemark{a} &  0\tablenotemark{a} & 18\tablenotemark{a} & 0.8$\pm$0.4 & 5\tablenotemark{b} & \nodata & 18\tablenotemark{b} & $<$2.6\tablenotemark{c}\\
 & 6\tablenotemark{a} & 18.0$\pm$4.7 & 18\tablenotemark{a} & 0.6$\pm$0.1 & 5\tablenotemark{a} & 25.4$\pm$4.0 & 18\tablenotemark{a} & 1.2$\pm$0.5 & 5\tablenotemark{b} & \nodata & 18\tablenotemark{b} & $<$2.6\tablenotemark{c}\\
  & 6\tablenotemark{a} & 41.1$\pm$6.2 & 18\tablenotemark{a} & 0.3$\pm$0.1 & 5\tablenotemark{b} & \nodata & 18\tablenotemark{b} & $<$1.0\tablenotemark{c} & 5\tablenotemark{b} & \nodata & 18\tablenotemark{b} & $<$2.6\tablenotemark{c}\\
  & 6\tablenotemark{a} & 64.4$\pm$4.7 & 18\tablenotemark{a} & 0.3$\pm$0.1 & 5\tablenotemark{b} & \nodata & 18\tablenotemark{b} & $<$0.8\tablenotemark{c} & 5\tablenotemark{b} & \nodata & 18\tablenotemark{b} & $<$2.6\tablenotemark{c}\\
\enddata
\tablenotetext{a}{Parameter fixed in the LTE analysis.}
\tablenotetext{b}{Assumed parameter to derive an upper limit on the molecular column density.}
\tablenotetext{c}{3$\sigma$ upper limit as the molecular emission is not detected at the velocity shown by the HOC$^+$ emission.}
\end{deluxetable*}
\end{longrotatetable}

\begin{figure*}
\caption{HCO F=2-1, 1-0, 1-1, 0-1, 4-3, 3-2 (J=7/2-5/2), 3-2 (J=5/2-3/2), 2-1 and 2-2 transitions (labeled in the panel of position 13) observed towards thirteen positions of Sgr B2 (shown in Fig.~\ref{Pos_figure}). The positions are labeled with numbers on the left top corner of each panel. The LTE best fits to the lines obtained with MADCUBA are shown with red lines. Blue lines indicate the frequencies of nine HCO transitions. Other identified and unidentified spectral features are shown in panels of positions 1 and 3. The red line indicates where the 3 mm band ends.}
\centering
\includegraphics[width=0.83\textwidth]{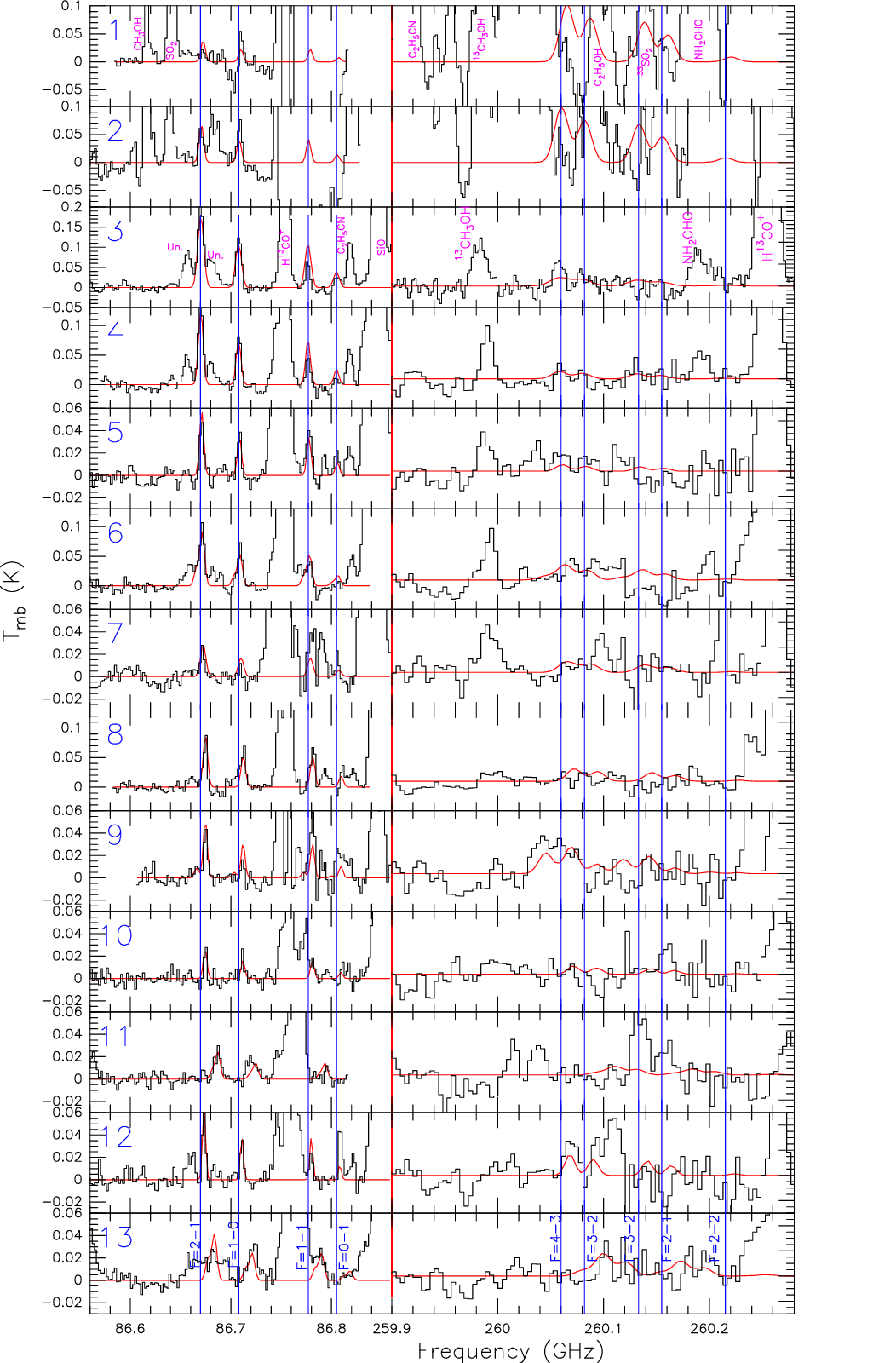}
\label{hco_figure}
\end{figure*}

\begin{figure*}
\caption{CO$^+$ transitions observed towards thirteen positions (shown in Fig.~\ref{Pos_figure}) of Sgr B2. The positions are labeled with numbers on the left top corner of each panel. Red lines show the overall spectral fitting to the lines, considering the contribution from $^{13}$CH$_3$OH in positions 1-3 and from (CH$_3$)$_2$CO in positions 1-2. Blue lines show the contribution of the CO$^+$ lines to the overall spectral fitting in positions 1-3. Synthetic spectrum of $^{13}$CH$_3$OH is shown in positions 4 and 6 with the aim of ruling out the presence of CO$^+$ emission. The positions of the four brighter transitions of $^{13}$CH$_3$OH J=5-4 are indicated with vertical red lines in positions 1-4 and 6, as well as those of (CH$_3$)$_2$CO in positions 1 and 2. Our best fit to the CO$^+$ lines are shown with blue lines in position 9. The fitting to the lines is explained in Section \ref{COp_section}. Dashed lines represent the velocity components shown by the C$^{18}$O J=1-0 lines (see Figure~\ref{co_figure}) towards the same positions.}
\includegraphics[width=1.03\textwidth]{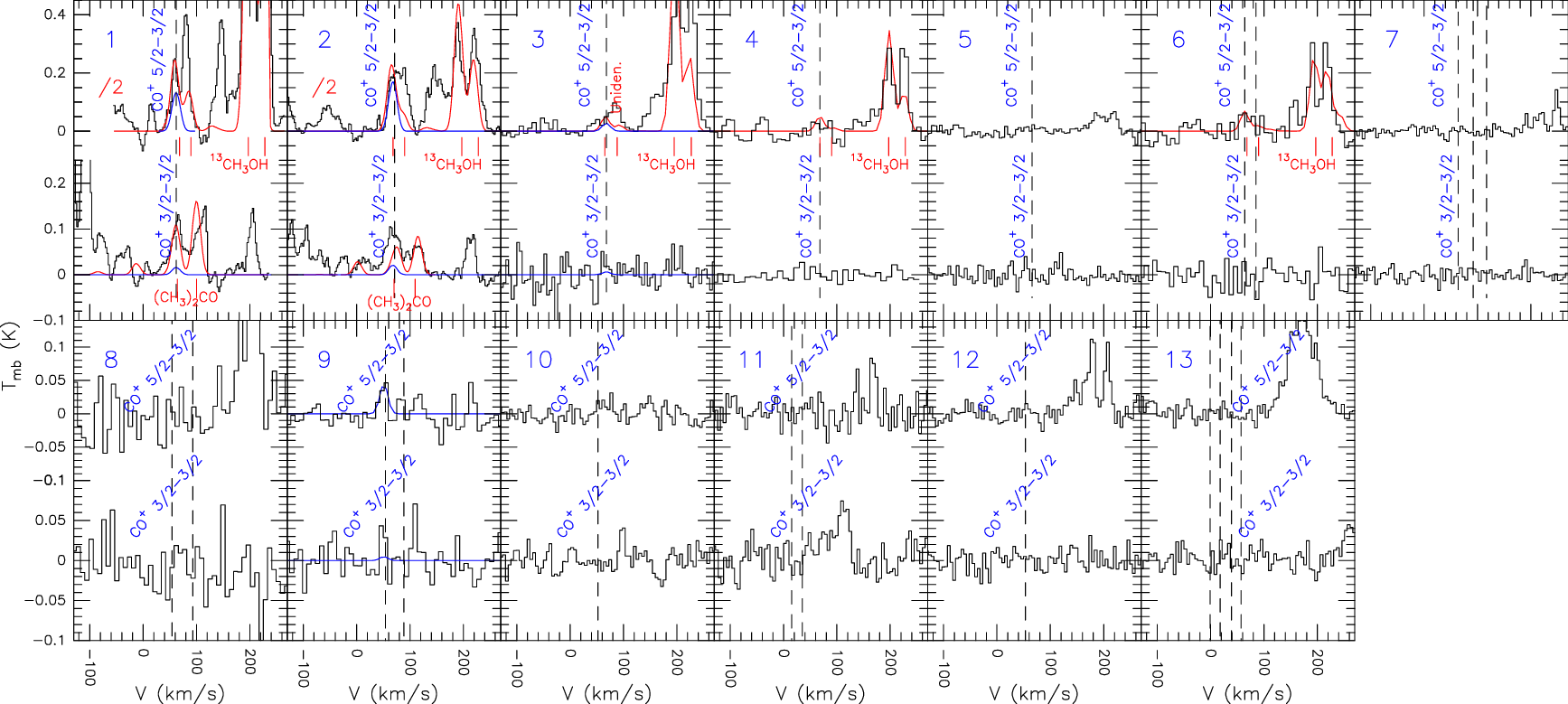}
\label{cop_figure}
\end{figure*}

\begin{figure*}
\caption{C$^{18}$O J=1-0 lines observed towards thirteen positions (shown in Fig.~\ref{Pos_figure}) of Sgr B2. The positions are labeled with numbers on the left top corner of each panel. The intensity of the C$^{18}$O J=1-0 spectrum is divided by 3 towards positions 1 and 2 for better visualization. The LTE best fits to the lines obtained with MADCUBA are shown with red lines. Dashed lines represent the velocity components found from the line fitting. Multiple velocity components are identified towards positions 6-9, 11 and 13 (see Section \ref{C18O_section}).}
\includegraphics[width=1\textwidth]{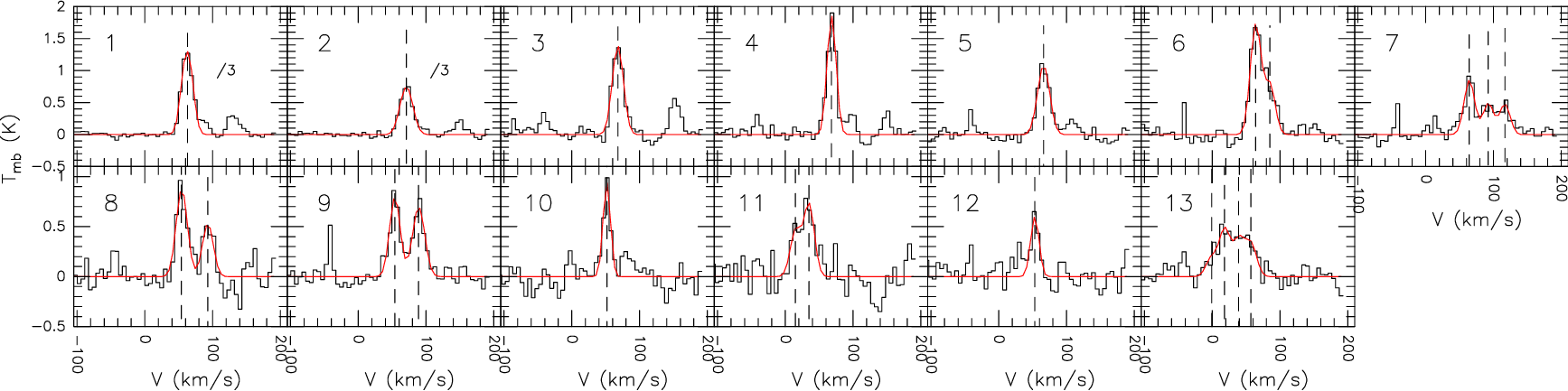}
\label{co_figure}
\end{figure*}

\begin{figure*}
\caption{HC$^{18}$O$^+$ J=1-0 and 3-2 lines observed towards thirteen positions of Sgr B2. The positions are labeled with numbers on the left top corner of each panel. The intensity of the J=1-0 and 3-2 spectra of HC$^{18}$O$^+$ is divided by 5 and 12, respectively, towards positions 1 and 2 for better visualization. The LTE best fits to the lines obtained with MADCUBA are shown with red lines. Dashed lines represent the velocity components found from the line fitting. Multiple velocity components are identified towards positions 6--9, 11 and 13 (see Section \ref{hc18op_section}). Other spectral features are labeled in the spectra of position 1.}
\centering
\includegraphics[width=1\textwidth]{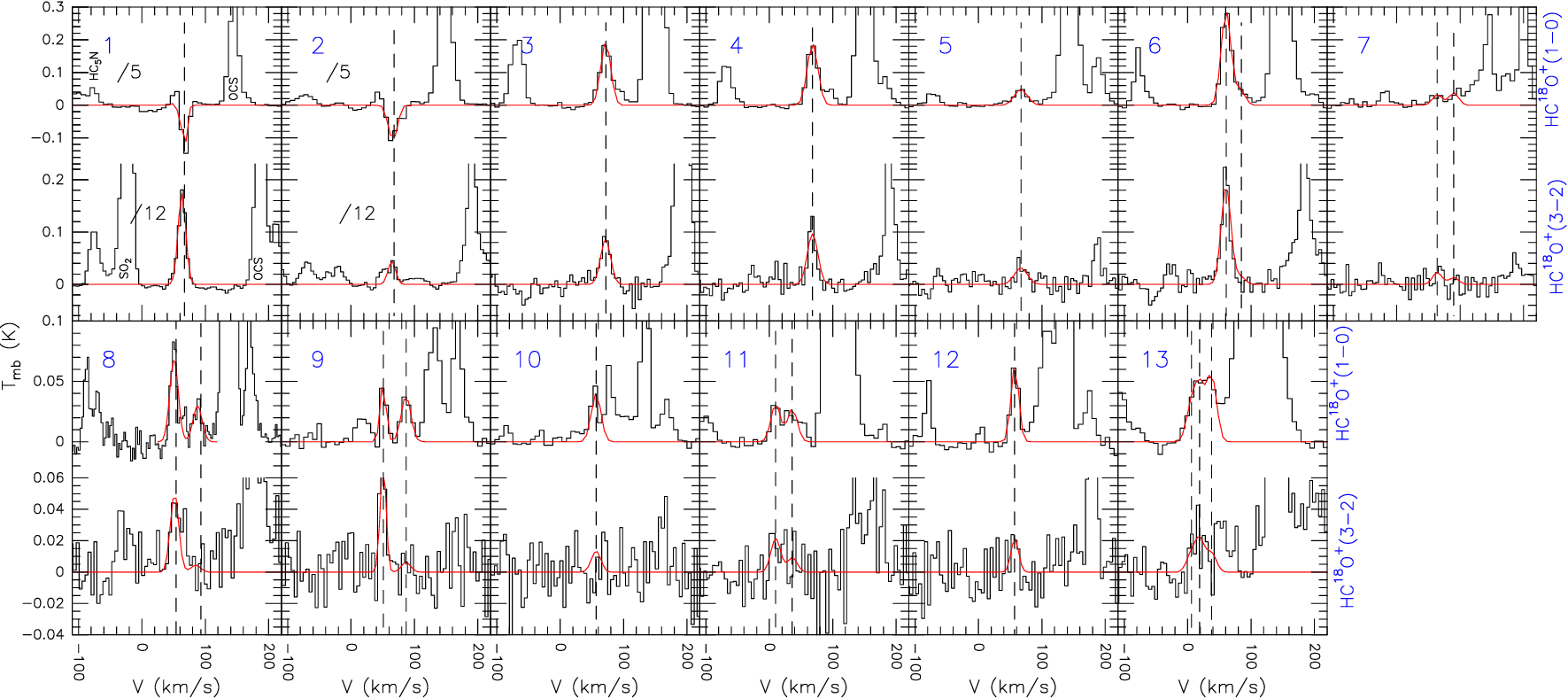}
\label{hc18o_figure}
\end{figure*}

In our study, we also used On-the Fly (OTF) data observed on August of 2006 (Program 012-06; PI: Mart\'in-Pintado). With this observing mode we mapped a region of 15'$\times$14' within the Sgr B2 complex. The data were calibrated using the standard dual load system. The A100 and B100 receivers connected to the 1 MHz filter bank were used in our observations, covering the frequencies of the SiO J=2-1, C$^{18}$O J=1-0, H$^{13}$CO$^+$ J=1-0 and HCO 1$_{0,1}$-0$_{0,0}$ (J=3/2-1/2, F=2-1) transitions. Our OTF observations of HCO 1$_{0,1}$-0$_{0,0}$ cover a region of 9'$\times$9', which is smaller than that mapped in the other molecular lines (see Figures~\ref{Pos_figure} and \ref{hcoemission}).
The data reduction was done with the GILDAS software\footnote{http://www.iram.fr/IRAMFR/GILDAS}, which was also used to build the SiO J=2-1, H$^{13}$CO$^+$ J=1-0 and HCO 1$_{0,1}$-0$_{0,0}$ data cubes with half-power beams of $\sim$28 arcsec.
The data cubes of SiO, H$^{13}$CO$^+$, C$^{18}$O and HCO have an rms noise level of $\sim$25 mK in the T$_{\rm a}^*$ scale ($\sim$30 mK in T$_{\rm mb}$).
Contours tracing the SiO J=2-1 emission integrated in two velocity ranges are shown in Figure~\ref{Pos_figure}. 
Integrated intensity maps of the HCO 1$_{0,1}$-0$_{0,0}$ and H$^{13}$CO$^+$ J=1-0 lines are shown in Figure~\ref{hcoemission}.
The C$^{18}$O J=1-0 spectra extracted over 22 arcsec regions at selected positions will be used later to derive molecular abundances.

\subsection{XMM-Newton observations and data reduction}
We have used observations available in the XMM-Newton archive to study the X-ray emission from Sgr B2. The selected data were observed with the EPIC/MOS instruments in 2004 (ObsID:0203930101; PI: Decourchelle) and 2012 (ObsID:0694640601;PI: Terrier). The data taken in 2004 and 2012 have exposure times of 50918 and 41921 seconds, respectively.
The emchain task of the Science Analysis Software (SAS, version 16.1.0) was used to run the default pipeline processing, as well as to obtain the calibrated event lists, which were cleaned for flaring events. Then, the evselect task of SAS was used to extract the two images in the range of 6.3-6.5 keV shown in Figure~\ref{Pos_figure}. This energy range is chosen with the aim of mapping the 6.4 keV Fe line emission and the underlying continuum from Sgr B2.

\begin{deluxetable*}{c|cccc|cccc}
\tablecaption{Derived parameters from the C$^{18}$O J=1-0 and HC$^{18}$O$^+$ J=1-0 lines towards thirteen positions of Sgr B2. For deriving the column density N$_{\rm tot}$ of C$^{18}$O the excitation temperature T$_{\rm ex}$ is assumed to be equal to 10 K for all positions and velocity components (see text).\label{C18O_parameters}}
\tablehead{
Pos. & \multicolumn{4}{c}{C$^{18}$O} & \multicolumn{4}{c}{HC$^{18}$O$^+$} \\
 & V$_{\rm LSR}$ & FWHM & N$_{\rm tot}$ & N$_{\rm H_2}$ & T$_{\rm ex}$ & V$_{\rm LSR}$ & FWHM &  N$_{\rm tot}$\\
     & (km s$^{-1}$) & (km s$^{-1}$) & ($\times$10$^{16}$ cm$^{-2}$) &($\times$10$^{22}$ cm$^{-2}$) & (K) & (km s$^{-1}$) & (km s$^{-1}$) & ($\times$10$^{12}$ cm$^{-2}$)}
\startdata
1  & 62.4$\pm$0.5 & 18.6$\pm$1.2 & 12.0$\pm$0.8 &31.0$\pm$2.0 & 1\tablenotemark{a} & 66\tablenotemark{a} & 12\tablenotemark{a} & 10.0$\pm$3.2\\
2  & 70.9$\pm$0.5 & 21.0$\pm$1.0 & 6.8$\pm$0.3 &17.0$\pm$0.7 & 1\tablenotemark{a} & 65\tablenotemark{a} & 12\tablenotemark{a} & 7.1$\pm$1.6\\
3  & 67.8$\pm$0.5 & 18.8$\pm$1.1 & 3.4$\pm$0.2 &8.5$\pm$0.4 & 6.3$\pm$0.3 & 71.7$\pm$0.6 & 18\tablenotemark{a} & 6.0$\pm$0.4\\
4 & 68.3$\pm$0.6 & 14.9$\pm$1.3 & 3.7$\pm$0.3 &9.3$\pm$0.7 & 6.7$\pm$0.4 & 67.9$\pm$0.7 & 18.0$\pm$1.7 & 6.0$\pm$0.6\\
5 & 66.1$\pm$0.7 & 20.6$\pm$1.7 & 2.8$\pm$0.2 &7.0$\pm$0.5 & 7.3$\pm$0.7 & 67.1$\pm$1.5 & 21.3$\pm$3.6 & 1.8$\pm$0.3\\
6 & 64.1$\pm$0.5 & 18\tablenotemark{a} & 4.2$\pm$0.2 &10.0$\pm$0.5 & 7.2$\pm$0.2 & 60.6$\pm$0.4 & 18\tablenotemark{a} & 8.9$\pm$0.5\\
  & 85.1$\pm$1.2 & 18\tablenotemark{a} & 1.7$\pm$0.2 & 4.2$\pm$0.4 & 7\tablenotemark{a} & 85.1$\pm$3.1 & 18\tablenotemark{a} & 1.3$\pm$0.3\\
7 & 64.3$\pm$0.9 & 18\tablenotemark{a} & 1.9$\pm$0.2 &4.8$\pm$0.4 & 7.6$\pm$1.7 & 65.0$\pm$3.0 & 18\tablenotemark{a} & 1.0$\pm$0.4\\
  & 92.3$\pm$1.7 & 18\tablenotemark{a} & 1.0$\pm$0.2 &2.6$\pm$0.4 & 7\tablenotemark{a} & 94\tablenotemark{a} & 18\tablenotemark{a} & 1.2$\pm$0.5\\
  & 116.7$\pm$1.7 & 18\tablenotemark{a} & 1.0$\pm$0.2 &2.5$\pm$0.4 & \nodata & \nodata & \nodata & \nodata\\
8 & 54.4$\pm$1.0 & 19.1$\pm$2.4 & 2.0$\pm$0.2 &5.1$\pm$0.6 & 8.7$\pm$1.1 & 53.4$\pm$1.4 & 18.8$\pm$3.7 & 1.6$\pm$0.3\\
  & 92.9$\pm$1.7 & 18.7$\pm$4.0 & 1.2$\pm$0.2 &2.9$\pm$0.5 & 7\tablenotemark{a} & 85\tablenotemark{a} & 18\tablenotemark{a} & 0.6$\pm$0.2\\
9 & 53.8$\pm$0.7 & 18.1$\pm$1.6 & 1.8$\pm$0.1 & 4.4$\pm$0.3 & 10.4$\pm$1.4 & 50.5$\pm$1.1 & 11.0$\pm$1.6 & 0.9$\pm$0.2 \\
  & 89.0$\pm$0.8 & 20.7$\pm$2.0 & 1.8$\pm$0.1 &4.4$\pm$0.4 & 4.6$\pm$2.5 & 87.1$\pm$4.6 & 18\tablenotemark{a} & 1.4$\pm$1.0\\
10 & 51.6$\pm$0.9 & 12.9$\pm$2.0 & 1.5$\pm$0.2 &3.7$\pm$0.5 & 7\tablenotemark{a} & 53\tablenotemark{a} & 15\tablenotemark{a} & 1.3$\pm$0.4\\
11 & 15.3$\pm$2.1 & 18\tablenotemark{a} & 1.6$\pm$0.2 & 2.5$\pm$0.4 & 7.3$\pm$1.2 & 9.3$\pm$2.3 & 18\tablenotemark{a} & 0.9$\pm$0.2\\
   & 35.2$\pm$1.4 & 18\tablenotemark{a} & 1.0$\pm$0.2 & 4.0$\pm$0.5 & 5.8$\pm$1.4 & 36.0$\pm$3.0 & 18\tablenotemark{a} & 0.9$\pm$0.3\\
12 & 53.1$\pm$1.7 & 14.4$\pm$3.9 & 1.0$\pm$0.2 &2.6$\pm$0.6 & 5.7$\pm$0.6 & 57.1$\pm$0.9 & 13.9$\pm$2.2 & 1.5$\pm$0.2\\
13 & -0.5$\pm$4.0 & 18\tablenotemark{a} & 0.4$\pm$0.1 &1.1$\pm$0.4 & 6.6$\pm$1.2 & 11.6$\pm$2.5 & 18\tablenotemark{a} & 1.4$\pm$0.6\\
& 18.3$\pm$2.4 & 18\tablenotemark{a} & 1.0$\pm$0.1 &2.6$\pm$0.4 & 5.1$\pm$1.8 & 27.8$\pm$3.4 & 18\tablenotemark{a} & 1.2$\pm$0.3\\
& 39.1$\pm$2.9 & 18\tablenotemark{a} & 0.8$\pm$0.1 &2.0$\pm$0.4 & 5.6$\pm$3.6 & 39\tablenotemark{a} & 18\tablenotemark{a} & 1.5$\pm$0.6\\
& 57.1$\pm$2.5 & 18\tablenotemark{a} & 0.7$\pm$0.2 &1.8$\pm$0.4 & \nodata & \nodata & \nodata & \nodata \\
\enddata
\tablenotetext{a}{Parameter fixed in the LTE analysis.}
\end{deluxetable*}

\section{Analysis and results of the X-ray data}\label{Xrays}
The X-ray emission in the 6.3-6-5 keV range is shown in Figure~\ref{Pos_figure}, where contours show the SiO J=2-1 emission in the velocity ranges of [65-75] and [15-25] km s$^{-1}$ in the middle and bottom panels, respectively. The SiO J=2-1 emission in the range of [65-75] km s$^{-1}$ matches better the 2004 X-ray features towards the Sgr B2M (position 1) and Sgr B2N (position 2) hot cores, while the SiO J=2-1 emission in the range of [15-25] km s$^{-1}$ coincides well with the brightest feature in the 2012 X-ray map.
The X-ray map observed in 2004 shows several emission sources (G0.66-0.03, G0.74-0.10 and G0.66-0.13) studied in \cite{Terrier18}. The selected thirteen positions observed with the IRAM 30 meter telescope are indicated on the X-ray maps in Figure~\ref{Pos_figure}. As mentioned, positions 1-12 were selected for our study as their chemistry seem to be affected by UV photons and/or X-rays, while position 13 traces quiescent gas.
G0.66-0.03 and G0.74-0.10 in the 2004 X-ray map match several positions observed with the IRAM 30 meter telescope.
G0.66-0.03, associated with the Sgr B2 core \citep{Terrier18}, match our observed positions 1-8, while the location of G0.74-0.10 overlaps with position 12 (see Figure~\ref{Pos_figure}). As seen in Figure~\ref{Pos_figure}, positions 10 and 11 also overlap with the X-ray emission. All these positions overlap with the X-ray emission above 6$\sigma$. 
There is a third large X-ray feature, that has not been labeled before, located just north of position 11, which we name as G0.69-0.11. On the other hand, G0.66-0.13 on the 2012 X-ray map is located just south of position 11.
G0.66-0.13 seems to be associated with the southeast region of the shell shown by the SiO J=2-1 emission at 15-25 km s$^{-1}$ (see Figure~\ref{Pos_figure}). The variability in the X-ray emission observed in Sgr B2 has been explained as a result of an echo from a limited number of relatively short flares likely originated in Sgr A* \citep{Terrier18}.

We have extracted X-ray spectra towards the thirteen positions (over regions with 29\arcsec in diameter, matching thus the highest HPBW of our IRAM 30m observations) of Sgr B2 to measure the Fe K$\alpha$ line flux shown in Figure~\ref{Fe6.4_flux}. An upper limit to the Fe K$\alpha$ line flux is given in this figure when the Fe line is not detected. The Fe K$\alpha$ emission line is not detected in positions 3, 5, 7-11 and 13 in the 2004 data, as well as in none of the thirteen positions in the 2012 data. Figure~\ref{Xray_spectrum} shows a sample of the X-ray spectra observed in 2004 towards positions 1, 3, 9 and 12.

We have modeled the X-ray emission arising from positions 1, 3, 9 and 12 considering models that include a hot thermal emission added to a spectrum produced by summing a power law and one Gaussian for the Fe K$\alpha$ line (in the case that it is detected, otherwise the Gaussian is omitted in the modeling). Line of sight absorption is also considered in the emission modeling. The best-fit models of the four positions are shown in Figure~\ref{Xray_spectrum} and the derived parameters are listed in Table \ref{Xray_table}. Following the work by \cite{Terrier10}, the intrinsic absorption and plasma temperature are assumed to be equal to 6.8$\times$10$^{23}$ cm$^{-2}$ and 5.8 keV, respectively. Both parameters together with the solar metallicity were considered as fixed parameters in our modeling.

Table \ref{Xray_table} shows that the studied regions would be irradiated by a field of X-rays lower than 10$^{37.09}$ erg s$^{-1}$. Positions 1, 2, 4, 6 and 12 (see Figure~\ref{Fe6.4_flux}) show a decrease in the Fe K$\alpha$ flux larger than $\sim$60\% between 2004 and 2012. 
Assuming a constant decay in the X-ray flux from 2004 to 2012, the 2012 fluxes have decreased an extra 15\% until 2014. Regions studied towards positions 6 and 12 can be considered as source targets to study the possible effects of X-rays on the molecular gas because they do not trace ionized gas (see Section \ref{ionized_hydrogen} and Figure~\ref{h42a_figure}), but they do trace gas that has been affected by X-rays in 2004, which could have modified the chemistry of the gas towards both regions. We have also derived X-ray fluxes within $\sim$10$^{-14}$-10$^{-13}$ erg s$^{-1}$ cm$^{-2}$ in the 2-10 keV band for all positions using our models. The possible effects of the X-rays on the chemistry of the molecular gas will be discussed in Section \ref{X_ray_effects}.

\section{Analysis and results of the mm data}\label{Analysis}
\subsection{Maps of HCO and H$^{13}$CO$^+$}
As mentioned above, Figure~\ref{hcoemission} displays maps of the HCO 1$_{0,1}$-0$_{0,0}$ (F=2-1) and H$^{13}$CO$^+$ J=1-0 lines integrated over the velocity range of 40-80 km s$^{-1}$. As seen in this figure, the HCO 1$_{0,1}$-0$_{0,0}$ line emission is extended and detected above 2$\sigma$ towards positions 3-4 and partially in position 6.
As mentioned in Section \ref{observa}, the HCO data cube has an rms level of $\sim$30 mK. This rms value is not enough to detect the HCO 1$_{0,1}$-0$_{0,0}$ F=2-1 line emission (above 2$\sigma$) towards positions 1, 2, 5, 7-13 in our map because in Section \ref{HCO_section} we will notice that this HCO line has peak intensities $\lesssim$60 mK towards these ten positions.

Figure~\ref{hcoemission} shows widespread  H$^{13}$CO$^+$ J=1-0 line emission above 3$\sigma$ towards Sgr B2.
Figure~\ref{Pos_SgrB2} displays widespread SiO J=2-1 line emission towards Sgr B2 as well. As seen in Figures \ref{Pos_figure} and \ref{hcoemission}, the molecular emission towards Sgr B2 is extended over the beam of the 30 meter IRAM telescope. Thus, in Sections \ref{hocp_section}-\ref{hc18op_section} we have considered a beam-filling factor of unity in the line modeling.

\subsection{Ionized hydrogen}\label{ionized_hydrogen}
As seen in Figure~\ref{h42a_figure}, we have detected the line emission $\gtrsim$55 mK from the H42$\alpha$ radio recombination line (RRL) towards positions 1-3, 5, 7-9 in Sgr B2. The HII regions 1-2, 5, 7-9 were selected from 20 cm continuum emission peaks shown in Figure~\ref{Pos_figure}. Following the same method as in Section 3.4 from \cite{Armijos18}, we have derived electron densities of 280-550 cm$^{-3}$ for positions 3, 5, 7-9, as well as electron densities of 1200-1700 cm$^{-3}$ for positions 1 and 2. We have considered the sources in positions 3, 5, 7-9 as diffuse HII regions as their electron densities are at least a factor 2.2 lower than those of positions 1 and 2 considered as dense HII regions. Positions 9 has the highest electron density of 540 cm$^{-3}$ among all positions tracing diffuse HII gas.

There is no emission of the H42$\alpha$ RRL in positions 11-13 and it is very low in positions 4, 6 and 10, supporting the idea that these sources are quiescent affected by shocks, and some of them affected by X-rays (see Section \ref{Xrays}). Towards these quiescent regions we estimated electron densities lower than 180 cm$^{-3}$. Figure~\ref{h42a_figure} reveals that the ionized gas does not have a molecular counterpart in all velocity components in positions 7-9. As it will be discussed below, the C$^{18}$O gas has two velocity components at $\sim$54 and 90 km s$^{-1}$ in positions 8 and 9, while the ionized gas shows only low velocities within $\sim$55-70 km s$^{-1}$ in both positions (see Figure~\ref{h42a_figure}). In position 7, the C$^{18}$O gas shows three velocity components at 
$\sim$64, 92 and 117 km s$^{-1}$, whereas the H42$\alpha$ RRL intensity peaks at 90-120 km s$^{-1}$.

\subsection{HOC$^+$ molecule}\label{hocp_section}
Figure~\ref{hocp_figure} shows spectra of the HOC$^+$ 1-0 and 3-2 transitions towards the thirteen positions studied in this paper. The HOC$^+$ 1-0 line is observed in absorption in position 1 and appears slightly absorbed in position 2. The HOC$^+$ 3-2 line is blended with hyperfine lines of $^{33}$SO$_2$(7$_{2,6}$-6$_{1,5}$). The SO$_2$ molecule is considered as a good tracer of hot core chemistry \citep{Jimenez07}. As seen in Figure~\ref{hocp_figure}, the HOC$^+$ J=1-0 emission is detected in positions 3-13, whereas the HOC$^+$ J=3-2 emission above 3$\sigma$ is detected only in positions 3, 4, 6 and 9.

To derive HOC$^+$ column densities (N$_{\rm tot}$) and excitation temperatures, we have used the AUTOFIT tool of SLIM in the MADCUBA package \citep{Martin19}. This tool allows us to fit synthetic spectra to the data assuming Local Thermodynamic Equilibrium (LTE) conditions.
The full widh at half maximum (FWHM) of the HOC$^+$ lines was fixed in the line fitting (except for the low velocity gas of position 9). Multiple velocity components were considered in the line fitting of positions 6-9, 11 and 13. The excitation temperature (T$_{\rm ex}$) in the range from 4.6 to 9.9 K is found towards several positions studied in Sgr B2 (see Table \ref{Tab_column}). The T$_{\rm ex}$ was fixed to the average value of 6 K when the AUTOFIT tool did not converge.

Sgr B2M and Sgr B2N were considered as blackbody emitters with a size of 2$\arcsec$ and a temperature of 150 K \citep{Belloche13} to fit the absorption line of HOC$^+$ J=1-0 in positions 1 and 2, respectively. This continuum was also used in \cite{Rivilla18} to study both sources.
As mentioned above, the HOC$^+$ 3-2 emission is blended with the $^{33}$SO$_2$ emission.
\cite{Belloche13} explained the $^{33}$SO$_2$ emission in Sgr B2M (our position 1) using two components. One considers a source size of 2$\arcsec$, a T$_{\rm ex}$ of 200 K and a column density of 1.8$\times$10$^{17}$ cm$^{-2}$, while the other considers a source size of 60$\arcsec$, a T$_{\rm ex}$ of 50 K and a column density of 5.4$\times$10$^{13}$ cm$^{-2}$.
The observed HOC$^+$ J=3-2 line in position 1 was modeled using both the $^{33}$SO$_2$ emission from the compact component estimated by \cite{Belloche13} and the contribution of HOC$^+$ J=3-2. 
In this modeling, we have neglected the $^{33}$SO$_2$ emission contribution from the extended component as it represents $\sim$5\% of the observed HOC$^+$ J=3-2 line. An additional component for HOC$^+$ with a T$_{\rm ex}$ of 30 K and a column density of $\sim$1$\times$10$^{13}$ cm$^{-2}$ is needed to model the HOC$^+$ J=3-2 line in position 2, in which we have not taken into account a contribution from $^{33}$SO$_2$ as this species has not been found in position 2 \citep{Belloche13}.
The derived parameters of HOC$^+$ are listed in Table \ref{Tab_column}.

\subsection{HCO molecule}\label{HCO_section}
Figure~\ref{hco_figure} displays spectra at 1 and 3 mm covering four hyperfine transitions (F=2-1, 1-0, 1-1, 0-1) and five hyperfine transitions (F=4-3, 3-2, 3-2, 2-1, 2-2) of HCO, respectively.
The HCO F=2-1 and 1-0 transitions are detected in positions 1 and 2, whereas the HCO F=1-1 and 0-1 transitions are blended with the strong absorption of the H$^{13}$CO$^+$ J=1-0 transition in both positions. The emission from the HCO F=2-1, 1-0 and 1-1 transitions is detected in positions 3-13, while that of the HCO F=0-1 transition is not detected in all positions. 
The F=2-1 transition of HCO shows peak intensities $\lesssim$60 mK towards positions 1, 2, 5, 7-13 (see Figure~\ref{hco_figure}).
The emission from the 1 mm HCO transitions is not observed above 3$\sigma$ in position 3-13, except that of the HCO F=4-3 hyperfine transition in position 3. Spectral features of other molecules are also seen in Figure~\ref{hco_figure}, which are labeled in the 1 and 3 mm spectral windows of positions 1 and 3.

As in the case of the HOC$^+$ molecule, the AUTOFIT tool was used to estimate HCO column densities and excitation temperatures assuming LTE conditions. The HCO lines in positions 1 and 2 are modeled considering the same assumptions for the background sources as in the case of HOC$^+$. HCO shows an average T$_{\rm ex}$ \mbox{of $\sim$5 K} derived from seven positions of Sgr B2. A warmer component of HCO with a T$_{\rm ex}$ of 20 K and a column density of 2$\times$10$^{14}$ cm$^{-2}$ is needed to fit the low intensity lines of HCO in positions 1 and 2. The T$_{\rm ex}$, FWHM and the local standard of rest (LSR) velocity were fixed in modeling when the AUTOFIT tool did not converge in our modeling. Two velocity components are included in the line modeling of positions 6, 11 and 13. The LTE best fits to the HCO lines are shown in Figure~\ref{hco_figure}, and the derived parameters are listed in Table \ref{Tab_column}. Upper limits to the column density are given in this table when there is undetected HCO emission at a given velocity.

\subsection{CO$^+$ molecule}\label{COp_section}
CO$^+$ 5/2-3/2 and 3/2-3/2 spectra towards our studied positions of Sgr B2 are shown in Figure~\ref{cop_figure}. Unfortunately, the CO$^+$ 5/2-3/2 transition is blended with two hyperfine transitions of $^{13}$CH$_3$OH J=5-4, while the CO$^+$ 3/2-3/2 transition is blended with two transitions (12$_{11,1}$-11$_{10,2}$ and 12$_{11,2}$-11$_{10,1}$) of (CH$_3$)$_2$CO. The CO$^+$ lines are detected in positions 1-3 and 9, although without ambiguity only in position 9. The ambiguity in positions 1-3 is due to the line blending (see below).

We have used the AUTOFIT tool of MADCUBA to derive the CO$^+$ column density. The line fitting of the CO$^+$ lines in positions 1 and 2 is complex because of the line blending. The emission contribution from $^{13}$CH$_3$OH and (CH$_3$)$_3$CO to the CO$^+$ lines can not be ruled out in positions 1 and 2. For this reason the lines of the three molecular species are fitted simultaneously with the AUTOFIT tool. The T$_{\rm ex}$ of 10 K for $^{13}$CH$_3$OH and (CH$_3$)$_3$CO gives the best fits to the observed lines, which yields values of $\sim$2$\times$10$^{15}$ cm$^{-2}$ and $\sim$2$\times$10$^{16}$ cm$^{-2}$ for $^{13}$CH$_3$OH and (CH$_3$)$_3$CO, respectively. Several parameters were fixed in the CO$^+$ line fitting in all positions (see Table \ref{Tab_column}).
The LTE best fits to the observed CO$^+$ lines are given in Figure~\ref{cop_figure}, where the synthetic spectrum of $^{13}$CH$_3$OH (and also of (CH$_3$)$_2$CO for positions 1 and 2) is shown for positions 1-4 and 6 to evidence the emission contribution of $^{13}$CH$_3$OH to the CO$^+$ line. The synthetic spectrum of $^{13}$CH$_3$OH is obtained considering a T$_{\rm ex}$ of 10 K for the five positions. The estimated parameters are given in Table \ref{Tab_column}, where an upper limit to the column density is listed in case of undetected CO$^+$ lines.

\subsection{C$^{18}$O molecule}\label{C18O_section}
C$^{18}$O J=1-0 spectra towards the target positions of Sgr B2 are shown in Figure~\ref{co_figure}. The ground state line of C$^{18}$O is included in our work as a tracer of H$_2$, allowing us to derive relative abundances of HOC$^+$, HCO and CO$^+$. The C$^{18}$O J=1-0 lines were fitted using MADCUBA and considering multiple velocity components in several positions (see Figure~\ref{co_figure}). Using C$^{18}$O data, \cite{Martin08} derived the average T$_{\rm ex}$ of 10 K for several positions of Sgr B2, which is considered in the C$^{18}$O J=1-0 line fitting shown in Figure~\ref{co_figure}. Thus, we have derived the C$^{18}$O column density listed in Table \ref{C18O_parameters}. The H$_2$ column density (N$_{\rm H_2}$) is also given in this table, which was calculated from the column density of C$^{18}$O using the $^{16}$O/$^{18}$O isotopic ratio of 250 \citep{Wilson94} and the relative abundance of CO to H$_2$ of 10$^{-4}$ \citep{Frerking82}.

\subsection{HC$^{18}$O$^+$ molecule}\label{hc18op_section}
To estimate ratios of HCO$^+$ to HOC$^+$, HCO and CO$^+$, we have also analyzed the HC$^{18}$O$^+$ J=1-0 and 3-2 spectra shown in Figure~\ref{hc18o_figure}, towards our target positions in Sgr B2. As seen in this figure, the HC$^{18}$O$^+$ J=1-0 line is detected in all positions but not the HC$^{18}$O$^+$ J=3-2 line. The ground state transition of HC$^{18}$O$^+$ is absorbed in positions 1 and 2.

As in the case of the other molecules, we have fitted synthetic spectra of HC$^{18}$O$^+$ J=1-0 and 3-2 to our data to estimate the HC$^{18}$O$^+$ column density. The HC$^{18}$O$^+$ lines of positions 1 and 2 were modeled considering that Sgr B2M and Sgr B2N have continuum background sources emitting as backbodies with similar conditions to those used for HOC$^+$. An additional warmer component of HC$^{18}$O$^+$ with a T$_{\rm ex}$ of 21 K and a column density of (0.5-1)$\times$10$^{13}$ cm$^{-2}$ is needed to fit the HC$^{18}$O$^+$ lines in positions 1 and 2. Multiple velocity components are taken into account in the HC$^{18}$O$^+$ line fitting of positions 6-9, 11 and 13. The found parameters are listed in Table \ref{C18O_parameters}. As seen from this table, the temperature within $\sim$6-10 K explains the excitation of HC$^{18}$O$^+$ towards several positions studied in Sgr B2, except the uncertain T$_{\rm ex}$ of 1 K required to model the absorbed HC$^{18}$O$^+$ J=1-0 lines in positions 1 and 2.

\begin{figure*}
\caption{Histogram showing HOC$^+$ (in cyan), HCO (in red) CO$^+$ (in green) and HCO$^+$ (in magenta) abundances towards the thirteen Sgr B2 positions. More than one bar is used in positions 6-9, 11 and 13, where multiple velocity components of gas are found. In these cases, bars ordered from left to right correspond to gas components ranging from low to high velocities. The gas velocities in these cases can be checked for HOC$^+$, HCO and CO$^+$ in Table \ref{Tab_column} and for HC$^{18}$O$^+$ (used to derive the HCO$^+$ abundances, see Section \ref{hc18op_section}) in Table \ref{C18O_parameters}. The high velocity gas of positions 8 and 9 is indeed quiescent gas. Arrows above several bars indicate that these bars represent upper limits on the molecular abundances.}
\includegraphics[clip,trim=1.7cm 0 2cm 0,width=1.0\textwidth]{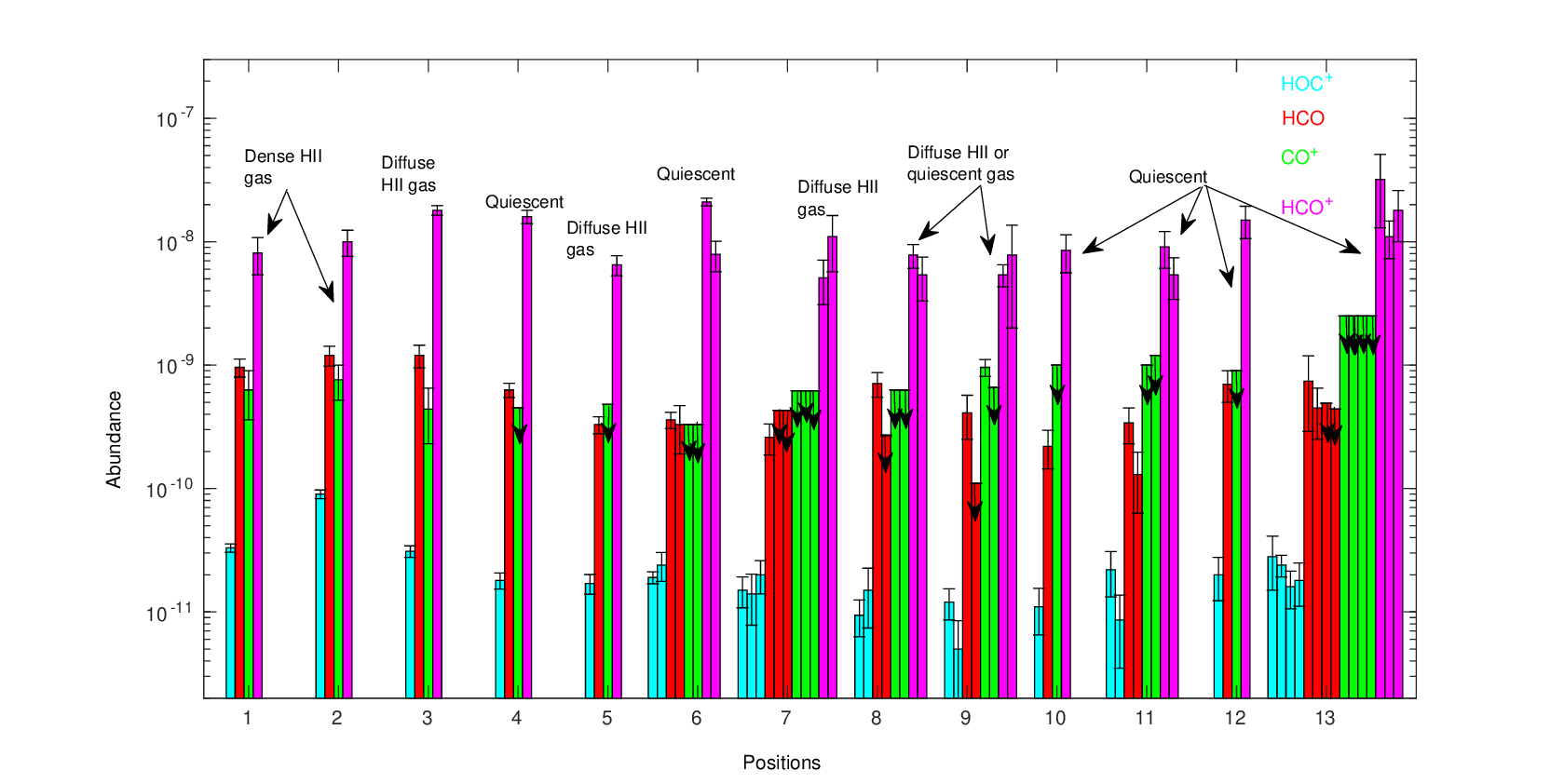}
\label{histogram}
\end{figure*}

In Section \ref{Ratios_section} we will use HCO$^+$ column densities derived from the HC$^{18}$O$^+$ column densities given in Table \ref{C18O_parameters} and using the $^{16}$O/$^{18}$O isotopic ration of 250 \citep{Wilson94}. From these HCO$^+$ column densities we have also derived relative abundances of HCO$^+$ within (0.5-3.2)$\times$10$^{-8}$ (see Figure~\ref{histogram}) considering the H$_2$ column densities listed in Table \ref{C18O_parameters}.

\section{Discussion}
\subsection{X-ray effects on the molecular gas?}\label{X_ray_effects}
As mentioned above, the variability in the X-ray emission in Sgr B2 could have been produced by several flares originated in Sgr A* \citep{Terrier10,Terrier18}. In this scenario, Sgr B2 and other GC clouds would act as reflection clouds. Figure~\ref{Pos_figure} shows that the gas mainly towards positions 1, 2, 6 and 12 was affected by X-ray emission in 2004, which may have affected the chemistry of the molecular gas. To address the possible effects of the X-rays on the molecular gas we have calculated the X-ray ionization rate $\zeta_X$ for Sgr B2 using the equation 4 given in \cite{Maloney96}:

\begin{equation}
\zeta_{\rm X}\sim1.4\times10^{-11} s^{-1} \left(\frac{L}{10^{44}\,{\rm erg\,s^{-1}}}\right)
\left(\frac{r}{10^2\,{\rm pc}} \right)^{-2} \left(\frac{N_{\rm H_2}}{10^{22}\, {\rm cm^{-2}}} \right)^{-1}
\end{equation}

where $L$ is the X-ray luminosity, $r$ is the distance to the X-ray source and $N_{\rm H_2}$ is the H$_2$ column density. We have derived the $\zeta_X\sim$10$^{-19}$ s$^{-1}$ considering $L\sim$10$^{37}$ erg s$^{-1}$ found for positions 1 and 12 in Section \ref{Xrays}, r=120 pc (the projected distance between Sgr B2 and Sgr A*) and the average column density N$_{\rm H_2}$ of 1$\times$10$^{23}$ cm$^{-2}$ estimated for positions 1, 2, 6 and 12 (see Table \ref{C18O_parameters}), where X-ray emission is detected in our 2004 map shown in Figure \ref{Pos_figure}. The $\zeta_{\rm X}$ value of $\sim$10$^{-19}$ s$^{-1}$ derived for Sgr B2 agrees with those shown in Figure 4 of \cite{Harada15} for the H$_2$ column densities higher than 10$^{23}$ cm$^{-2}$.
The effects of the X-rays on the chemistry of the molecular gas are similar to those of cosmic rays \citep{Harada15}. The cosmic-ray ionization rate $\zeta$ is (4-7)$\times$10$^{-16}$ s$^{-1}$ in Sgr B2 \citep{vanderTak06,Bonfand19}, which is three orders of magnitude higher than the $\zeta_{\rm X}$ derived for Sgr B2. The very low $\zeta_{\rm X}$ of $\sim$10$^{-19}$ s$^{-1}$ shows that the effects of the X-rays on the Sgr B2 molecular gas might be negligible.

\subsection{Molecular abundances}\label{Abundances}
Figure~\ref{histogram} shows the abundances of HOC$^+$, HCO, CO$^+$ and HCO$^+$ estimated for the thirteen positions of Sgr B2. There is more than one bar in positions 6-9, 11 and 13 in this figure because of the multiple velocity components. The gas in positions 1-8 and 10-12 was pervaded by X-rays in 2004 as mentioned in Sections \ref{sec:intro} and \ref{Xrays}, but from now the gas in positions 4, 10, the high velocity gas in positions 8 and 9, and the gas in positions 10-12 is considered quiescent as the effects of the X-rays on the Sgr B2 molecular gas might be negligible as shown in Section \ref{X_ray_effects}.
HOC$^+$ shows the highest abundance of 9$\times$10$^{-11}$ towards position 2 tracing dense HII gas.
The low velocity quiescent gas of position 13 shows also a high abundance of HOC$^+$, which is a factor of 3.2 lower than that in position 2. This suggests that HOC$^+$ is formed efficiently in regions affected by UV photons, but also through other chemical routes discussed below.

\begin{figure*}
\caption{Fractional abundances (top panels) and ratios (bottom panels) of observed molecules obtained with our chemical modeling for an H$_{\rm 2}$ density of 10$^4$ cm$^{-3}$, kinetic temperatures of 50 K (solid lines) and 100 K (dotted lines) and a visual extinction of 10 mag. The cosmic-ray ionization rate $\zeta$ ranges from 1$\times$10$^{-17}$ s$^{-1}$ to 1$\times$10$^{-14}$ s$^{-1}$.
On the top panels, we show with gray bands the HCO$^+$ abundances of (0.5-3.2)$\times$10$^{-8}$ (for the thirteen positions), the HOC$^+$ abundances of (0.5-2.8)$\times$10$^{-11}$ (for quiescent gas), the HCO abundances of (1-12)$\times$10$^{-10}$ (for diffuse HII and quiescent gas) and the CO$^+$ abundances of (4-10)$\times$10$^{-10}$ (for dense and diffuse HII gas) estimated in our study (see Section \ref{Abundances}). On the bottom panels, we show with gray bands the HCO$^+$/HOC$^+$ ratios of 337-1541 (for quiescent gas), the HCO$^+$/HCO of 11-20 and 24-60 for diffuse HII and quiescent gas, respectively, and the HCO$^+$/CO$^+$ ratios of 6-41 (for dense and diffuse HII gas) derived in Section \ref{Ratios_section}.}
\gridline{\includegraphics[clip,trim=2cm 0 0 0,width=1.0\textwidth]{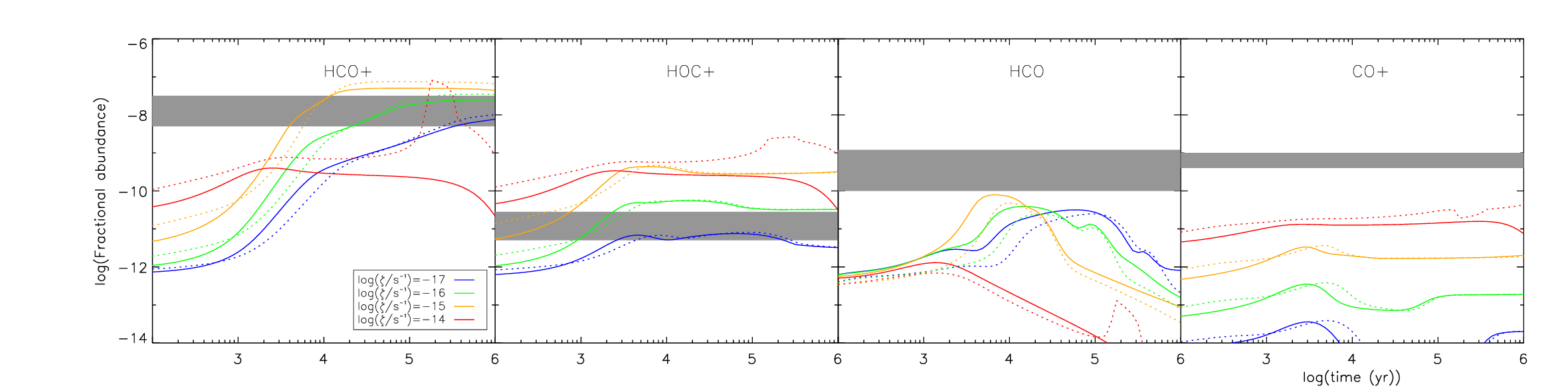}}
\gridline{\includegraphics[clip,trim=2.1cm 0 0 0.5cm,width=1.0\textwidth]{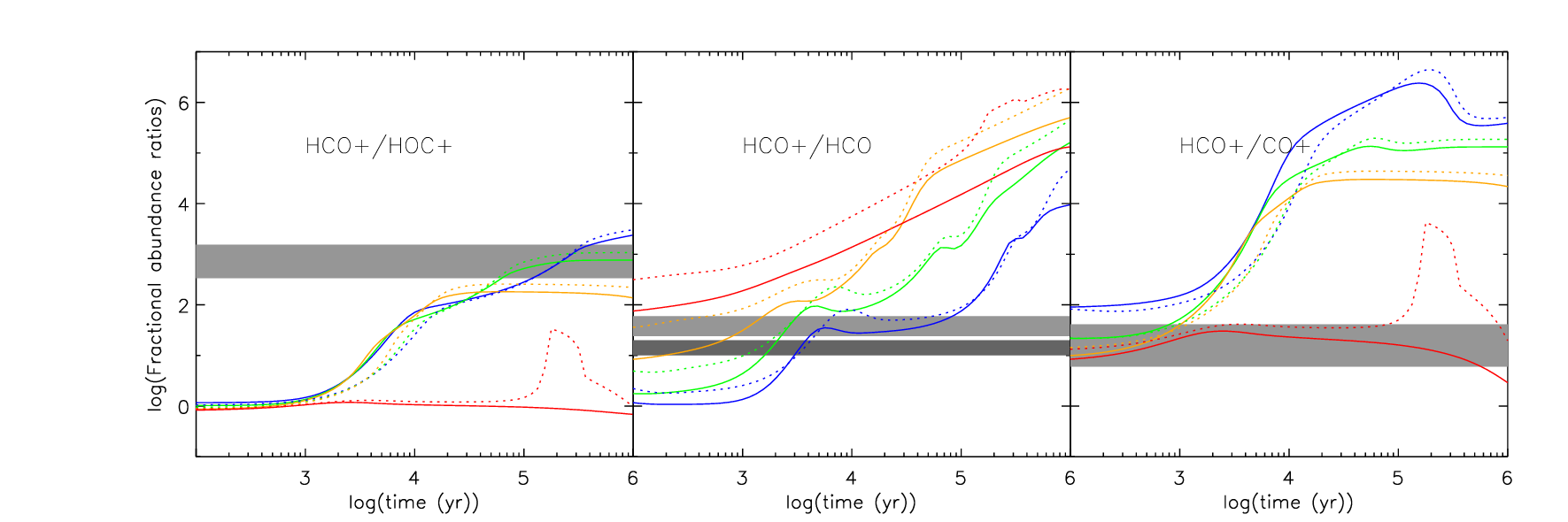}}
\label{Molecular_ratios}
\end{figure*}

Figure~\ref{Molecular_ratios} shows fractional abundances of HOC$^+$, HCO, and CO$^+$ obtained with the gas-grain time-dependent code Nautilus \citep{Hersant09,Semenov10}. The initial condition for hydrogen is in the molecular form, while for other elements with higher ionization potential than 13.6 eV  is in atomic form, and for all other elements is in the ionic form. We ran the model with constant physical conditions. The modeled abundances are obtained for an H$_2$ density of 10$^4$ cm$^{-3}$ found towards Sgr B2 \citep{Hutte95,Armijos15}, kinetic temperatures of 50 and 100 K, a visual extinction of 10 mag (implying H$_2$ column densities of $\sim$10$^{22}$ cm$^{-2}$, values estimated for several Sgr B2 positions as shown in Table \ref{C18O_parameters}) and the $\zeta$ values of 10$^{-17}$, 10$^{-16}$, 10$^{-15}$ and 10$^{-14}$ s$^{-1}$. $\zeta$ values of (4-7)$\times$10$^{-16}$ s$^{-1}$ are found towards Sgr B2 \citep{vanderTak06,Bonfand19}.
Kinetic temperatures of 50 and 100 K are included in our chemical modeling because these temperatures are the approximate lower and upper values of the range of kinetic temperatures found for GC molecular clouds, which may be driven by turbulent dissipation and/or cosmic rays \citep{Ao13}.
However, \cite{Ginsburg16} proposed that turbulent heating is the main heating mechanism of the GC dense gas. \cite{Meijerink11} predicted gas temperatures of $\sim$100-200 K for hydrogen column densities $>$10$^{22}$ cm$^{-2}$ (as those found for Sgr B2) and $\zeta$ values within 5$\times$10$^{-17}$-5$\times$10$^{-13}$ s$^{-1}$ using PDR models that mimic the effects of cosmic rays and mechanical heating on the gas, but when mechanical heating is not included in the modeling gas temperatures $>$50 K (for hydrogen column densities $>$10$^{22}$ cm$^{-2}$) are only reached for the high $\zeta$ values of 5$\times$10$^{-14}$ and 5$\times$10$^{-13}$ s$^{-1}$, which are at least a factor of 100 higher than that of $\sim$5$\times$10$^{-16}$ s$^{-1}$ estimated for Sgr B2 \citep{vanderTak06,Bonfand19}. 
This contrasts with what was found by \cite{Bisbas15,Bisbas17}, who predicted gas temperatures $\lesssim$50 K using models with $\zeta$ values $\leq$10$^{-14}$ s$^{-1}$.
Taking into account the $\zeta$ value of $\sim$5$\times$10$^{-16}$ s$^{-1}$ for Sgr B2, mechanical heating may be the mechanism responsible for increasing the gas temperatures to values $>$50 K.

As seen in Figure~\ref{Molecular_ratios}, the HOC$^+$ abundance increases as the value of the $\zeta$ increases. 
Figure~\ref{Molecular_ratios} shows differences of more than one order of magnitude in the HOC$^+$ abundances predicted by the models with the same $\zeta$ value of 10$^{-14}$ s$^{-1}$ but different kinetic temperatures of 50 and 100 K at timescales $>$10$^{5.0}$ years, whereas in the cases with the other $\zeta$ values these differences are of less than a factor of $\sim$3 over the timescales of 10$^2$-10$^6$ years.
HOC$^+$ abundances of $\sim$(0.5-2.8)$\times$10$^{-11}$ are estimated towards the positions of quiescent gas in Sgr B2.
Figure~\ref{Molecular_ratios} shows that the HOC$^+$ abundances of $\sim$(0.5-2.8)$\times$10$^{-11}$ are reached at timescales lower than $\sim$10$^{2.8}$ years for the $\zeta$ value of 10$^{-15}$ s$^{-1}$ or at timescales within $\sim$10$^{3.0}$-10$^{3.2}$ years for the $\zeta$ value of 10$^{-16}$ s$^{-1}$. Both $\zeta$ values are in agreement with those found for Sgr B2 \citep{vanderTak06,Bonfand19}.
The highest HOC$^+$ abundance of 2.8$\times$10$^{-11}$ found for the quiescent gas in Sgr B2 is very close to those predicted by models with the $\zeta$ value of 10$^{-16}$ s$^{-1}$ and timescales $>$10$^5$ years (see Figure~\ref{Molecular_ratios}).
Figure~\ref{Molecular_ratios} also shows HOC$^+$ abundances of $>$5$\times$10$^{-11}$ for the $\zeta$ value of 10$^{-14}$ s$^{-1}$ and different timescales. These HOC$^+$ abundances are at least a factor $\sim$1.8 higher than the highest HOC$^+$ abundance of 2.8$\times$10$^{-11}$ estimated for the quiescent gas in position 13.
Our derived abundances of HOC$^+$ are also consistent with those predicted by models with kinetic temperatures of 50 and 100 K, and the $\zeta$ value of 10$^{-17}$ s$^{-1}$ at timescales within $\sim$10$^{3.2}$-10$^{5.2}$ years, however the $\zeta$ value of 10$^{-17}$ s$^{-1}$ is at least a factor of 40 lower than those found in Sgr B2 \citep{vanderTak06,Bonfand19}.
This shows that the HOC$^+$ abundances observed in the Sgr B2 quiescent regions are well explained by a high temperature chemistry and the effects of cosmic rays with the $\zeta$ values of 10$^{-16}$ s$^{-1}$ and 10$^{-15}$ s$^{-1}$ at different timescales. Both $\zeta$ values will be considered in our discussion below.

The HOC$^+$ abundance of 4$\times$10$^{-9}$ is found towards the XDRs in the circumnuclear disk of NGC 1068 \citep{Usero04}, which is a factor of 143 higher than the highest HOC$^+$ abundance derived for the quiescent regions of Sgr B2, ruling out the effects of X-rays on the molecular gas and consistent with our discussion given in Section \ref{X_ray_effects}.

As seen in Figure~\ref{histogram}, the highest HCO abundances of $\sim$1$\times$10$^{-9}$ are found towards position 3 tracing diffuse HII gas, as well as towards positions 1-2, tracing dense HII gas. 
HCO shows also high abundances of $\sim$(3-7)$\times$10$^{-10}$ for the diffuse HII gas in positions 5, 7, 8 and 9, and of $\sim$(1-7)$\times$10$^{-10}$ for the quiescent gas in positions 4, 6, 10-13. This indicates that HCO forms efficiently in gas affected by UV photons but also in quiescent gas. It is believed that photo-processing of ice mantles and subsequent photodesorption of HCO or H$_2$CO (followed by gas phase photodissociation) are responsible for high HCO abundances of $\simeq$(1-2)$\times$10$^{-9}$ in PDRs \citep{Gerin09}, values that are consistent with those of $\sim$1$\times$10$^{-9}$ in positions 1-3.

Chemical modeling of a molecular cloud affected by non-dissociative shocks of 10 km s$^{-1}$ shows relative abundances of HCO higher than $\sim$1$\times$10$^{-9}$ starting at 10$^{3}$ years after the shock \citep{Mitchell84}, which is slightly higher than that of around 7$\times$10$^{-10}$ found for the low velocity quiescent gas in position 13.

Figure~\ref{Molecular_ratios} reveals differences of less than a factor of 10 between the HCO abundances predicted by models with the same $\zeta$ values of 10$^{-16}$ and 10$^{-15}$ s$^{-1}$ but different kinetic temperatures over the timescales of 10$^2$-10$^6$ years.
As seen in Figure~\ref{Molecular_ratios}, the highest HCO abundance of $\sim$8$\times$10$^{-11}$ is predicted for the $\zeta$ value of 10$^{-15}$ s$^{-1}$, the kinetic temperature of 50 K and timescales of $\sim$10$^{3.5}$ years. This HCO abundance is only a factor of 1.2 lower than that of 1$\times$10$^{-10}$ but it is a factor of 8.7 lower than the highest abundance of 7$\times$10$^{-10}$ found for the quiescent gas in position 13, suggesting that high temperature chemistry and cosmic rays are not enough to explain the highest HCO abundances observed in the Sgr B2 quiescent gas and other mechanisms such as shocks are needed to explain those high HCO abundances. It is believed that HCO in hot corinos is formed by the hydrogenation of CO on grains and subsequent thermal desorption in the protostellar phase \citep{Rivilla19}. This scenario is consistent with what we propose in our paper, with the difference that instead of desorption in the quiescent gas of Sgr B2 the HCO is released to the gas phase through grain sputtering by shocks.

On the other hand, CO$^+$ line emission is detected only in positions 1-3 and 9 (at gas velocities of $\sim$50 km s$^{-1}$), and without ambiguity only in position 9, as it was mentioned before. The CO$^+$ abundances in these four positions are within $\sim$(4-10)$\times$10$^{-10}$. We mentioned in Section \ref{ionized_hydrogen} that the H42$\alpha$ RRL has the highest electron density in position 9 between all the studied positions of diffuse HII gas. This suggests that CO$^+$ is produced mainly via reactions where UV photons play an important role. This is also supported by the coincidence between the CO$^+$ relative abundances derived in this paper and those of around 10$^{-9}$ shown by the PDR models of \cite{Martin09a}. Limits lower than $\sim$2$\times$10$^{-9}$ are estimated for diffuse HII gas in positions 5, 7-8 or quiescent gas in positions 4, 6, 10-13. The upper limits of these nine positions agree with those lower than $\sim$5$\times$10$^{-11}$ obtained for the four values of the $\zeta$ shown in Figure~\ref{Molecular_ratios}. This comparison points out that high abundances of CO$^+$ are expected only in PDRs as our chemical models including only high temperature chemistry and cosmic-ray effects show very low CO$^+$ abundances.

We found differences of less than a factor of 2 in the CO$^+$ abundances predicted by models with the same $\zeta$ values of 10$^{-16}$ and 10$^{-15}$ s$^{-1}$ but different kinetic temperatures of 50 and 100 K over timescales of 10$^2$-10$^6$ years (see Figure \ref{Molecular_ratios}).

\subsection{Column density ratios}\label{Ratios_section}
\begin{figure*}
\caption{\textbf{Top panels:} HCO$^+$/HOC$^+$ ratios are plotted for dense and diffuse HII regions, as well as for quiescent regions, as a function of the H$_2$ column density (left panel), the electron density (middle panel) and the fluxes of the 6.4 keV K$\alpha$ line (right panel). \textbf{Bottom panels:} same ratios as on the top panels but for the HCO$^+$/HCO ratio. Arrows drawn on several panels show lower or upper limits. The middle and right panels show less values than the left panel as the ionized gas does not have a molecular counterpart in all velocity components (see Section \ref{ionized_hydrogen}) and the fluxes of the 6.4 keV Fe line are represented only for the lowest velocity components in positions where multiple velocity components of gas are found.\label{Ratio_tracers}}
\gridline{\includegraphics[clip,trim=2.8cm 0 4cm 0,width=0.34\textwidth]{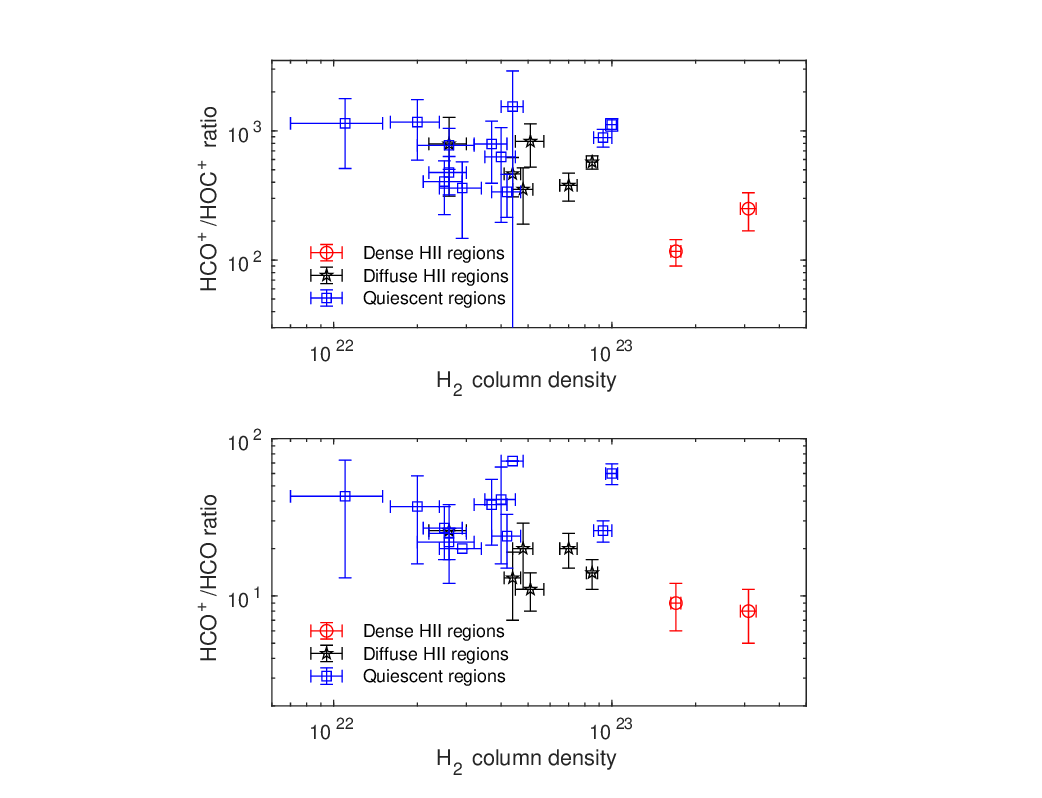}
\includegraphics[clip,trim=3.7cm 0 3.35cm 0,width=0.335\textwidth]{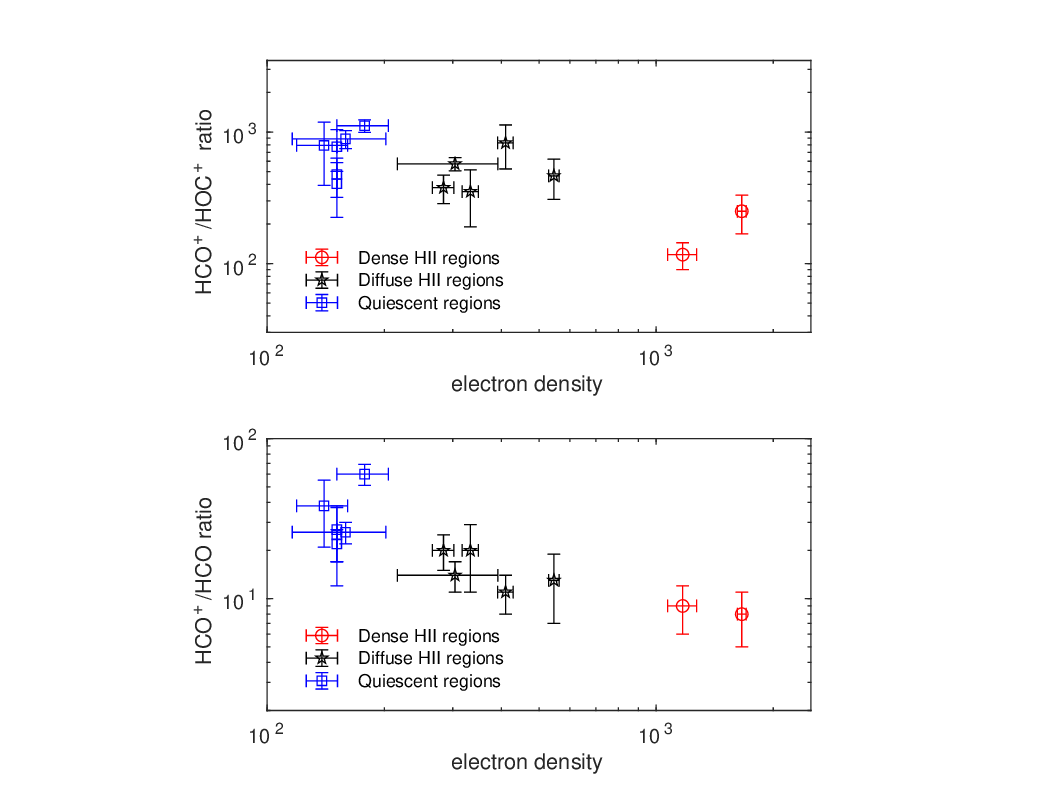}
\includegraphics[clip,trim=4.1cm 0 3.0cm 1.0cm,width=0.35\textwidth]{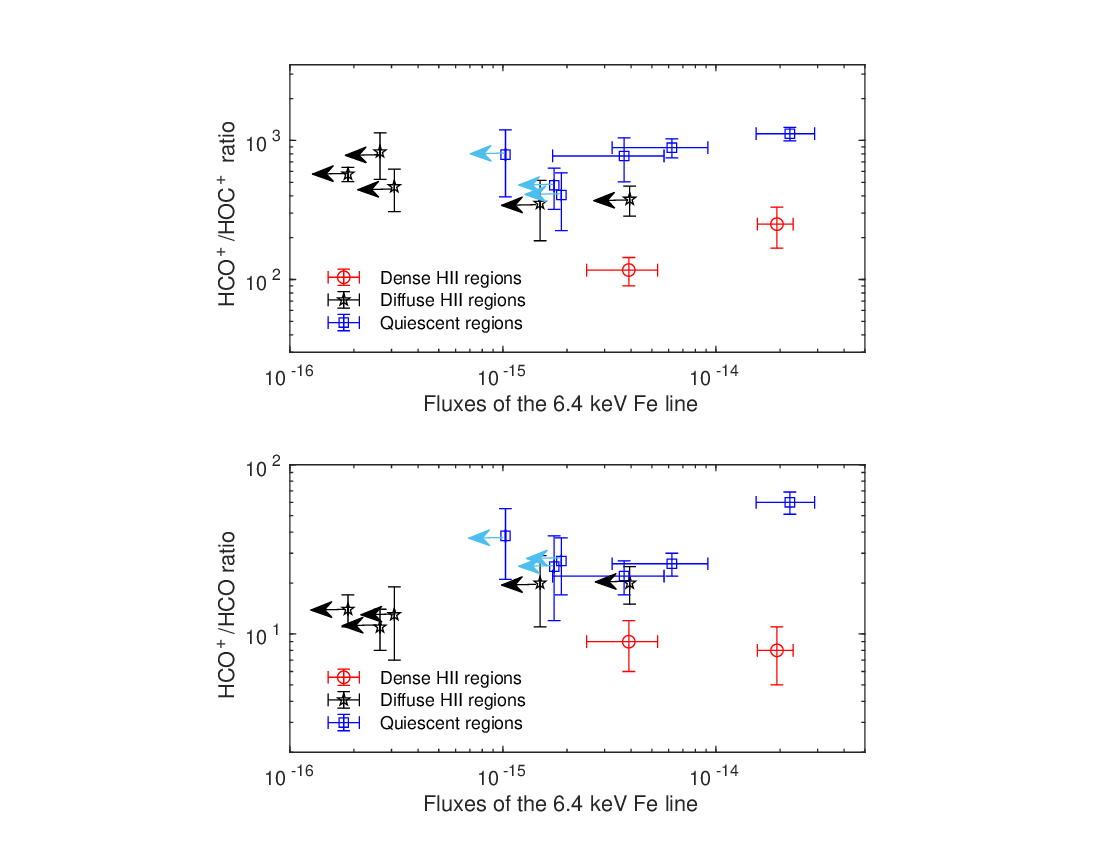}
         }
\end{figure*}

Column density ratios of HCO$^+$ versus HOC$^+$, HCO and CO$^+$ for the studied positions of Sgr B2 are given in Table \ref{ratios}. This table lists several velocity components for positions 6-9, 11 and 13 as multiple velocity components of gas are found (see Section \ref{Analysis}).
The ratios corresponding to NGC 253, M82, Horsehead and the Orion bar, considered as prototypical PDRs, are included in this table as well \citep{Martin09a,Goicoeche09,Gerin09,Fuente03,Apponi99,Schilke01,Savage04}. The HCO$^+$/HOC$^+$ and HCO$^+$/HCO ratios as a function of the H$_2$ column density, the electron density (derived in Section \ref{ionized_hydrogen}) and the fluxes of the 6.4 keV Fe line (derived from the 2004 X-ray data in Section \ref{Xrays}) for the positions studied in our paper are displayed in Figure~\ref{Ratio_tracers}. The lowest HCO$^+$/HOC$^+$ ratios of 117-250 are found in positions 1 and 2, while the other positions tracing diffuse HII regions or quiescent gas show HCO$^+$/HOC$^+$ ratios higher than 337. The bottom panels of Figure~\ref{Ratio_tracers} shows that there is a good correlation between the HCO$^+$/HCO ratio with both the H$_2$ column density and the electron density. Figure~\ref{Ratio_tracers} also shows that there is no correlation between the 6.4 keV Fe K$\alpha$ line flux and both the HCO$^+$/HOC$^+$ ratio and the HCO$^+$/HCO ratio.   

Figure~\ref{Ratio_tracers} shows similar values of the HCO$^+$/HOC$^+$ ratios both in diffuse HII regions and quiescent regions. 
The HCO$^+$/HOC$^+$ ratios of 117-250 agree with those of $<$270 of prototypical PDRs (see Table \ref{ratios}). Based on this, the HCO$^+$/HOC$^+$ ratio seems to be a good tracer of dense HII regions where UV photons dominate the heating and chemistry of the molecular gas, but it does not allow to distinguish between quiescent gas and diffuse ionized gas.

The bottom panels of Figure~\ref{Molecular_ratios} show the HCO$^+$/HOC$^+$, HCO$^+$/HCO and HCO$^+$/CO$^+$ ratios obtained with our chemical modeling. We believe that the agreement of the HCO$^+$/HOC$^+$ ratios of 337-1541 estimated for the Sgr B2 quiescent regions is better for the $\zeta$=10$^{-16}$ s$^{-1}$ as this $\zeta$ value agrees better with that of 4$\times$10$^{-16}$ s$^{-1}$ found for the Sgr B2 envelope and timescales $>$10$^{4.5}$ years. This is consistent with what was discussed in Section \ref{Abundances} and implying that the values of $\zeta$=10$^{-16}$ s$^{-1}$ and timescales $>$10$^{5.0}$ years give a better agreement to the observations. For these values our model reveals HCO$^+$ abundances of $\sim$1$\times$10$^{-8}$ that agrees with those of (0.5-3.2)$\times$10$^{-8}$ estimated for Sgr B2 (see Figure~\ref{Molecular_ratios}).

As seen in Figure~\ref{Ratio_tracers}, the HCO$^+$/HCO ratio also shows the lowest values within 8-9 in positions 1-2 tracing dense HII gas. On the other hand, the HCO$^+$/HCO ratios within 11-20 and 24-60 are found in diffuse HII gas and quiescent gas, respectively. It is clear in Figure~\ref{Ratio_tracers} that this ratio shows a range of higher values in quiescent regions than in regions of diffuse HII gas, implying that the HCO$^+$/HCO ratio is a good tracer of regions affected by UV photons as well as the HCO$^+$/HOC$^+$ ratio, but that this ratio is also sensitive to the electron density of the ionized gas. The HCO$^+$/HCO ratios of 24-60 found for quiescent gas agree with those estimated for the $\zeta$ value of 10$^{-15}$ s$^{-1}$, kinetic temperatures of 50 and 100 K, and timescales $<$10$^{3.1}$ years or the $\zeta$ value of 10$^{-16}$ s$^{-1}$, kinetic temperatures of 50 and 100 K, and timescales within 10$^{3.1}$-10$^{3.2}$ years. It is important to stress that the agreement of the modeled ratios for the above $\zeta$ and timescales with those derived is due to both the modeled HCO$^+$ and HCO abundances are the same orders of magnitude lower than those found for the quiescent gas (see Figure~\ref{Molecular_ratios}).

The HCO$^+$/HCO ratios within 8-9 derived for positions 1-2 are in agreement with those of the starburst galaxies NGC 253 and M 82 \citep{Martin09a}.

On the other hand, we were able to derive HCO$^+$/CO$^+$ ratios only for positions 1-3 and 9, where CO$^+$ emission was detected. We estimated lower limits on the HCO$^+$/CO$^+$ ratio for the other positions. As we can see in Table \ref{ratios}, the HCO$^+$/CO$^+$ ratio is also a good tracer of regions affected by UV photons as this ratio is within 6-14 in positions 1-2 and 9, where we found the highest electron densities between all the studied regions (see Section \ref{ionized_hydrogen}), while the HCO$^+$/CO$^+$ ratio of 41 is derived for the diffuse ionized gas in position 3. The HCO$^+$/CO$^+$ ratio of 41 found for position 3 is reached at the timescales lower than $\sim$10$^{3.0}$ years for the $\zeta$ values of 10$^{-16}$ s$^{-1}$ and 10$^{-15}$ s$^{-1}$ (see Figure~\ref{Molecular_ratios}). However, this agreement is due to both the modeled HCO$^+$ and CO$^+$ abundances are the same orders of magnitude lower than the derived values, as in the case of the HCO$^+$/HCO ratio.

\cite{Martin09a} found the HCO$^+$/CO$^+$ ratios of 32$\pm$16 and 38$\pm$15 for M 82 and NGC 253, respectively, which are between the values found for the dense and diffuse HII regions. The upper limit of $<$83-140 on the HCO$^+$/CO$^+$ ratio found for the Orion bar \citep{Fuente06,Savage04} is also consistent with those within 6-41 estimated towards positions 1-3 and 9.

\begin{deluxetable}{c|crrr}
\tablecaption{Ratios of HCO$^+$ versus HOC$^+$, HCO and CO$^+$\label{ratios}}
\tabletypesize{\scriptsize}
\tablehead{
 & & & & \\
Pos.(type of gas) & V$_{\rm LSR}$ & HCO$^+$/HOC$^+$ & HCO$^+$/HCO & HCO$^+$/CO$^+$\\
or  source    & (km s$^{-1}$) &  & &}
\startdata
1 (Dense HII gas) & 66 & 250$\pm$82 & 8$\pm$3 & 13$\pm$7\\
2 (Dense HII gas) & 65 & 117$\pm$27 & 9$\pm$3 & 14$\pm$5\\
3 (Diffuse HII gas) & 72 & 573$\pm$67 & 14$\pm$3& 41$\pm$20\\
4 (Quiescent gas) & 68 & 887$\pm$138 & 26$\pm$4 & $>$36\\
5 (Diffuse HII gas) & 67& 378$\pm$92 & 20$\pm$5 & $>$13\\
6 (Quiescent gas) & 61 & 1117$\pm$123 & 60$\pm$9 & $>$66\\
  (Quiescent gas) & 85 & 337$\pm$123 & 24$\pm$12 & $>$10\\
7 (Diffuse HII gas) & 65 & 353$\pm$163 & 20$\pm$9 & $>$8\\
   (Diffuse HII gas) & 94 & 791$\pm$478 & $>$26 & $>$10\\
8 (Diffuse HII gas) & 53 & 828$\pm$304 & 11$\pm$3 & $>$12\\
  (Quiescent gas) & 85 & 361$\pm$214 & $>$20 & $>$5\\
9 (Diffuse HII gas) & 50 & 465$\pm$157 &13$\pm$6 & 6$\pm$1\\
  (Quiescent gas) & 87 & 1541$\pm$1563 & $>$72 & $>$12\\
10 (Quiescent gas)& 53 &791$\pm$398 & 38$\pm$17 & $>$8\\
11 (Quiescent gas)& 9& 405$\pm$180 & 27$\pm$10 & $>$8\\
 (Quiescent gas) & 36& 628$\pm$432 & 41$\pm$25 & $>$6\\
12 (Quiescent)& 57& 773$\pm$270 & 22$\pm$5 & $>$17\\
13 (Quiescent gas)& 12& 1143$\pm$632 & 43$\pm$30 & $>$13\\
  (Quiescent gas) & 28&476$\pm$157 & 25$\pm$13 & $>$11\\
 (Quiescent gas) & 39& 1169$\pm$575 & 37$\pm$21 & $>$14\\
\hline
 & \multicolumn{4}{c}{Prototypical PDRs}\\
\hline
NGC 253 & \nodata  & 80$\pm$30\tablenotemark{a} & 5.2$\pm$1.8\tablenotemark{a} & 38$\pm$15\tablenotemark{a}\\
M 82 & \nodata & 60$\pm$28\tablenotemark{a} & 9.6$\pm$2.8\tablenotemark{a} & 32$\pm$16\tablenotemark{a}\\
Horsehead&\nodata & 75-200\tablenotemark{b} & 1.1\tablenotemark{c} & $>$1800\tablenotemark{b}\\
Orion bar & \nodata& $<$166-270\tablenotemark{d,e}& 2.4\tablenotemark{f} & $<$83-140\tablenotemark{d,g}\\
\enddata
\tablenotetext{a}{\cite{Martin09a}.}
\tablenotetext{b}{\cite{Goicoeche09}.}
\tablenotetext{c}{\cite{Gerin09}.}
\tablenotetext{d}{\cite{Fuente03}.}
\tablenotetext{e}{\cite{Apponi99}.}
\tablenotetext{f}{\cite{Schilke01}.}
\tablenotetext{g}{\cite{Savage04}.}
\end{deluxetable}

\section{Conclusions}
In this paper we compared SiO J=2-1 emission maps with X-ray maps observed in 2004 and 2012 towards Sgr B2. 
Integrated intensity maps of HCO 1$_{0,1}$-0$_{0,0}$ (J=3/2-1/2, F=2-1) and H$^{13}$CO$^+$ J=1-0 towards Sgr B2 are also presented.
Three large X-ray spots are observed in the X-ray map of 2004, two of them identified and studied by \cite{Terrier18}, while the other spot is identified in this paper and called as G0.69-0.11.
We also found that the known spot G0.66-0.13 observed in the 2012 X-ray map is apparently associated with a southeast region being part of a shell shown by the SiO J=2-1 emission at velocities of 15-25 km s$^{-1}$. 
In addition, we derived Fe K$\alpha$ line fluxes or limits on the those fluxes towards thirteen positions of Sgr B2. Twelve of these sources are affected by UV photons or/and X-rays, whereas the other remaining source is quiescent.
The molecular gas in these Sgr B2 positions was irradiated by a field of X-rays lower than 10$^{37.09}$ erg s$^{-1}$ until 2012. We derived the X-ray ionization rate of $\sim$10$^{-19}$ s$^{-1}$ for Sgr B2, a very low value to affect the chemistry of the molecular gas.

Our study of the H42$\alpha$ radio recombination lines and the derived electron density allowed us to classify the sources associated with HII regions into regions of diffuse and dense hydrogen ionized gas.
We found abundances of HOC$^+$, HCO and CO$^+$, as well as the HCO$^+$/HOC$^+$, HCO$^+$/HCO and HCO$^+$/CO$^+$ ratios for the thirteen positions, which were compared with those obtained with chemical modeling. We found that a model including a high temperature chemistry, cosmic rays with an ionization rate of 10$^{-16}$ s$^{-1}$ and timescales $>$10$^{5.0}$ years gives a better agreement of the HOC$^+$ abundances derived for quiescent gas of Sgr B2. Apart from high temperature chemistry and cosmic rays, shocks have to be invoked to explain the highest HCO abundances observed in the Sgr B2 quiescent gas. On the other hand, CO$^+$ emission is not detected towards quiescent regions, indicating that high abundances of CO$^+$ are expected only in PDRs.

We found differences of less than a factor of 3, 10 and 2 between the abundances of HOC$^+$, HCO and CO$^+$, respectively, predicted by models with the same cosmic-ray ionization rates of 10$^{-16}$ and 10$^{-15}$ s$^{-1}$ but different kinetic temperatures of 50 and 100 K over timescales of 10$^2$-10$^6$ years.

The lowest values of the HCO$^+$/HOC$^+$, HCO$^+$/HCO and HCO$^+$/CO$^+$ ratios are found in positions of Sgr B2 tracing dense HII gas, which shows that these ratios are good tracers of regions where UV photons dominate the heating and chemistry of the molecular gas. The HCO$^+$/HCO ratio is sensitive to the electron density of ionized regions as it shows values within 12-18 and 7-9 for diffuse and dense HII regions, respectively, whereas HCO$^+$/HCO ratios within 27-60 are estimated for quiescent gas of Sgr B2. We found a good correlation between the HCO$^+$/HCO ratio with both the H$_2$ column density and the electron density.



\acknowledgments
This work is based on observations carried out with the IRAM 30-m telescope. IRAM is supported by INSU/CNRS (France), MPG (Germany) and IGN (Spain). MLl acknowledges support from CONICYT- PFCHA/Doctorado Nacional/2019-21191036. V.M.R. has received funding from the European Union's Horizon 2020 research and innovation programme under the Marie Sk\l{}odowska-Curie grant agreement No 664931. We thank the anonymous referees for the useful comments that improved the manuscript.

\end{document}